\documentclass[11pt,twoside,a4paper]{article}

\usepackage{hyperref}
\usepackage{amsmath}
\usepackage{amssymb}
\usepackage[pdftex]{color,graphicx}
\usepackage[english]{babel}
\usepackage[latin9]{inputenc}
\usepackage[left=2.5cm,right=2.5cm,top=2.5cm,bottom=2.5cm]{geometry}
\usepackage{color}
\usepackage{multirow}
\usepackage{enumerate}
\usepackage{mdwlist}
\usepackage{verbatim}

\newcommand{\xunit}[1]{\mathrm{#1}}
\newcommand{\unit}[1]{\; \mathrm{#1}}
\newcommand{\keV}{\unit{keV}}
\newcommand{\GeV}{\unit{GeV}}
\newcommand{\TeV}{\unit{TeV}}
\newcommand{\xGeV}{\xunit{GeV}}
\newcommand{\kpc}{\unit{kpc}}
\newcommand{\xkpc}{\xunit{kpc}}
\newcommand{\km}{\unit{km}}
\newcommand{\m}{\unit{m}}
\newcommand{\cm}{\unit{cm}}
\newcommand{\xcm}{\xunit{cm}}
\newcommand{\sqcm}{\cm^{2}}
\newcommand{\sqkm}{\km^{2}}
\newcommand{\cubcm}{\cm^{3}}
\newcommand{\xcubcm}{\xcm^{3}}
\newcommand{\percubcm}{\cm^{-3}}
\newcommand{\xpercubcm}{\xcm^{-3}}
\newcommand{\persqkm}{\km^{-2}}
\newcommand{\kmpers}{\unit{km/s}}
\newcommand{\yr}{\unit{yr}}
\newcommand{\Gyr}{\unit{Gyr}}
\newcommand{\pers}{\unit{s}^{-1}}
\newcommand{\peryr}{\unit{yr}^{-1}}

\title{}
\date{}

\begin{document}
\vspace{1.cm}
\begin{flushright}
ULB-TH/09-02
\end{flushright}
\vspace{1.cm}

\begin{center}

{\Large {\bf  Neutrinos from Inert Doublet Dark Matter}}
\vspace{0.8cm}\\
{\large  Sarah Andreas$^{\dagger,\ddagger}$, Michel H.G.
Tytgat$^\dagger$ and Quentin Swillens$^{\dagger,\star}$}
\vspace{0.35cm}\\
{\small $^\dagger${\it Service de Physique Th\'eorique, Universit\'e Libre de Bruxelles,\\
CP225, Bld du Triomphe, B-1050 Brussels, Belgium}
\vspace{0.15cm}\\
$^\ddagger${\it Institut f\"ur Theoretische Physik E\\RWTH Aachen
University, D-52056 Aachen, Germany}
\vspace{0.15cm}\\
$^\star${\it IIHE-Inter-University Institute for High Energies, \\
Universit\'{e} Libre de Bruxelles, B-1050 Brussels, Belgium}
\vspace{0.15cm}\\
\emph{Sarah.Andreas@rwth-aachen.de; mtytgat@ulb.ac.be;
qswillen@ulb.ac.be} }

\vspace{0.5cm}

\end{center}

\begin{abstract}
We investigate the signatures of neutrinos produced in the
annihilation of WIMP dark matter in the Earth, the Sun and at the
Galactic centre within the framework of the Inert Doublet Model and
extensions. We consider a dark matter candidate, that we take to be
one of the neutral components of an extra Higgs doublet, in three
distinct mass ranges, which have all been shown previously to be
consistent with both WMAP abundance and direct detection experiments
exclusion limits. Specifically, we consider a light WIMP with mass
between 4 and 8 GeV (low), a WIMP with mass around 60-70 GeV
(middle) and a heavy WIMP with mass above 500 GeV (high). In the
first case, we show that capture in the Sun may be constrained using
Super-Kamiokande data. In the last two cases, we argue that indirect
detection through neutrinos is challenging but not altogether
excluded. For middle masses, we try to make the most benefit of the
proximity of the so-called 'iron resonance' that might enhance the
capture of the dark matter candidate by the Earth. The signal from
the Earth is further enhanced if light right-handed Majorana
neutrinos are introduced, in which case the scalar dark matter
candidate may annihilate into pairs of mono-energetic neutrinos. In
the case of high masses, detection of neutrinos from the Galactic
centre might be possible, provided the dark matter abundance is
substantially boosted.
\end{abstract}

\section{Introduction}

Cosmological observations concur to indicate that about 25 \% of the
energy density of the universe is made of dark matter
(DM)~\cite{Seljak:2006bg,Yao:2006px}. The simplest and most popular
paradigm postulates that dark matter is a weakly interacting massive
particle or WIMP~\cite{Jungman:1995df,Bertone:2004pz}. The reason is
twofold. Firstly, a stable particle with annihilation cross section
in the picobarn range, characteristic of weak interactions, would
have a relic density in agreement with observations, $\Omega_{DM}
\propto 1/\langle \sigma v \rangle \sim 0.25$. Secondly,
supersymmetric extensions of the Standard Model predict the
existence of a dark matter candidate which, generically, is a spin
$1/2$ particle with weak scale interactions, the neutralino. There
is also a popular spin $1$ WIMP candidate, in the form of the
lightest Kaluza-Klein partner of the photon in models with Universal
Extra Dimensions (UED)~\cite{Hooper:2007qk}. To complete the list,
one may consider the Inert Doublet Model (IDM), a very simple
extension of the Standard Model introduced in
ref.~\cite{Deshpande:1977rw}. This model has two Higgs doublets and,
to prevent FCNC, a discrete $Z_2$ symmetry. The dark matter
candidate is the lighest neutral component of the extra Higgs
doublet.\footnote{This gives us a complete family of WIMPs. One
should add the gravitino, a spin 3/2 state, and a spin 2 candidate
might also exist in theories of modified
gravity~\cite{Dubovsky:2004ud}, or as a Kaluza-Klein partner of the
graviton in theories with extra dimensions. These higher spin DM
candidates have essentially gravitational interactions and are not
WIMPs.} Clearly, it is not the ambition of the IDM to compete with,
say, the MSSM or UED models. This model is however very simple and,
nevertheless, it does have interesting phenomenological
consequences, some of which have been emphasized in recent works,
refs.~\cite{Ma:2006km,Barbieri:2006dq,Hambye:2007vf}. Furthermore,
its dark matter features are not less interesting, both from the
perspective of direct detection and indirect detection, be it
through gamma rays~\cite{LopezHonorez:2006gr,Gustafsson:2007pc},
neutrinos~\cite{Agrawal:2008xz}, or antimatter~\cite{Nezri:2009jd}.

To complete the analysis of IDM dark matter phenomenology initiated
in
refs.~\cite{Ma:2006km,Barbieri:2006dq,Hambye:2007vf,LopezHonorez:2006gr,Gustafsson:2007pc,Agrawal:2008xz,Nezri:2009jd,Majumdar:2006nt,Andreas:2008xy},
we discuss further the signatures of the IDM dark matter candidate
in neutrinos.\footnote{This is a further step toward filling the gap
in the table I given in ref.~\cite{Taoso:2007qk}.} Specifically, we
consider three distinct dark matter mass ranges which we call in the
sequel, the low, middle and high mass ranges.  All are consistent
with WMAP abundance and we consider the potential for indirect
detection of these IDM dark matter candidates through the flux of
neutrinos that might be produced in its annihilation in the Sun (low
mass range), in  the Earth (middle mass range), or at the Galactic
centre (GC) (high mass range).

As discussed in ref.~\cite{Andreas:2008xy}, the IDM dark matter
candidate may be rather light, $m_{DM} \sim$ few GeV (low mass
range). Because of its potentially large cross section on nuclei, it
has been advocated as one of the possible explanations for the
recent combined DAMA/NaI and DAMA/Libra results,
ref.~\cite{Andreas:2008xy}. Such a candidate may  be a  source of
gammas from the GC~\cite{Andreas:2008xy,Feng:2008dz} and  neutrinos
from the Sun~\cite{Savage:2008er,Feng:2008qn}. Here, we discuss the
constraint posed on the IDM candidate by the limits from the
Super-Kamiokande experiment.

However, the most natural dark matter candidate in the IDM has a
mass $m_{DM} \sim 60 \GeV$ (middle mass
range)~\cite{Barbieri:2006dq,LopezHonorez:2006gr}. Furthermore,
being a scalar, it has dominantly spin-independent interactions with
nuclei. These features open the interesting possibility to enhance
the capture of the IDM candidate by the Earth, through resonant
scattering on heavy nuclei, in particular iron. This point has been
already emphasized in~\cite{Agrawal:2008xz}, an article that
appeared while our work was being completed. Our results on
neutrinos and muons fluxes, obtained using different tools but
making similar approximations regarding capture in the Earth, are in
agreement. However, unlike in~\cite{Agrawal:2008xz}, we also take
into account the constraints from direct detection experiments that
lead to more pessimistic conclusions. In particular, we compare the
flux of neutrinos with the expected sensitivity of IceCube, a km$^3$
neutrino telescope under construction at the South Pole, and show
that this detector may at most give limits that are complementary,
but not better than those given by existing and forthcoming dark
matter direct detection experiments. In an attempt to improve this
situation, we then consider a scenario based on a natural extension
of the IDM with three generations of right-handed Majorana
neutrinos. We show that, provided we somewhat depart from the scheme
proposed in~\cite{Ma:2006km} for radiative SM neutrinos masses, some
scalar candidates may have large annihilation rates into
mono-energetic SM neutrinos, hence strongly boosting the possibility
of observing indirect detection of dark matter captured by the
Earth.

Finally, the IDM dark matter may be heavy, $m_{DM} \gtrsim 500 \GeV$
(high mass range)~\cite{LopezHonorez:2006gr}. In this case, it has a
very small cross section on nuclei and is therefore not likely to be
captured by the Sun or by the Earth. On the other hand, it may
accumulate at the centre of the Galaxy and its annihilation may
produce high energy neutrinos that could be observed on Earth.
However, we show that the flux is small for mundane astrophysical
assumptions regarding the abundance of dark matter at the Galactic
centre. Here, we set an upper bound on the neutrino flux that might
be produced by assuming that the annihilation of the IDM candidate
does not give gamma rays in excess of those observed by the EGRET
satellite.

This paper is organized as follow. The Inert Doublet Model is
introduced in section~\ref{sec-Model}. In
section~\ref{sec-abundance}, we fix our conventions on the dark
matter abundance in the Galaxy, the Sun and the Earth. The
section~\ref{sec-Capture} is devoted to a review of capture and
annihilation of WIMP dark matter in the Sun and in the Earth and
their possible signatures in neutrino detectors or telescopes. In
section~\ref{sec-NuFluxGC}, we envision the possibility of observing
neutrinos produced at the Galactic centre. The predictions of the
IDM are discussed in section~\ref{sec-IDMresults}, followed by our
conclusions.

\section{The Inert Doublet Model} \label{sec-Model}

The Inert Doublet Model is a two Higgs doublet model, $H_1$ and
$H_2$, with an unbroken $Z_2$ symmetry under which
$$
H_1 \rightarrow H_1 \;\; \mathrm{and} \;\; H_2\rightarrow - H_2
$$
and all the other Standard Model particles are
even~\cite{Ma:2006km,Barbieri:2006dq}. The potential of the IDM can
be written as
\begin{eqnarray}
\label{potential} V &=& \mu_1^2 \vert H_1\vert^2 + \mu_2^2 \vert
H_2\vert^2  + \lambda_1 \vert H_1\vert^4 + \lambda_2 \vert
H_2\vert^4 + \lambda_3 \vert H_1\vert^2 \vert H_2 \vert^2 \\
& & + \lambda_4 \vert H_1^\dagger H_2\vert^2 + \frac{\lambda_5}{2}
\left[(H_1^\dagger H_2)^2 + h.c.\right]. \nonumber
\end{eqnarray}
In this model, $H_1$ contains the standard Brout-Englert-Higgs
particle $h$ (the Higgs for short) and the discrete symmetry, which
prevents FCNC, gives a dark matter candidate in the form of one of
the neutral components of the extra doublet $H_2 = (H^+, 1/\sqrt{2}
(H_0 +i A_0))^T$. The couplings and masses of the scalar particles
are related by
\begin{eqnarray}
m_h^2 &=& - 2 \mu_1^2 \; \equiv \; 2 \lambda_1 v^2 \quad \mathrm{with} \quad v  = 246 \GeV \nonumber\\
m_{H^+}^2 &=& \mu_2^2 + \lambda_3 v^2/2\nonumber\\
m_{H_0}^2 &=& \mu_2^2 +  (\lambda_3 + \lambda_4 + \lambda_5) v^2/2\nonumber\\
m_{A_0}^2 &=& \mu_2^2 +  (\lambda_3 + \lambda_4 - \lambda_5) v^2/2
\, . \label{masses}
\end{eqnarray}
Depending on quartic couplings, either $H_0$ or $A_0$ can be the
lightest $Z_2$-odd particle. We choose $H_0$ and,
following~\cite{Barbieri:2006dq}, we define $\lambda_L=(\lambda_3 +
\lambda_4 + \lambda_5)/2$, the coupling between $h$ and a pair of
$H_0$. In our investigation of the model we choose $\mu_2$,
$\lambda_2$ and the masses of scalar particles, including the Higgs,
as input parameters.

Experimental constraints on the IDM model are discussed
in~\cite{Barbieri:2006dq} and further in~\cite{Lundstrom:2008ai}
and~\cite{Cao:2007rm}. The latter also discusses the prospect for
discovery of the $A_0$ and $H_0$ at the LHC.
In~\cite{Lundstrom:2008ai}, the LEP I and II results on the
neutralino are used to put constraints on the mass range of $H_0$
and $A_0$, see in particular their figure~8. In the present work, we
consider that $H_0$ is the dark matter candidate. The $A_0$ masses
(and that of $H^\pm$) we will use are always consistent with LEP
data.

Assuming that $H_0$ were in thermal equilibrium in the early
universe, there are essentially three distinct $H_0$ mass ranges
that are consistent with the WMAP abundance for dark matter. In the
sequel, we refer to them as the low mass ($3 \GeV \lesssim m_{H_0}
\lesssim 8 \GeV$), the middle mass (\mbox{$40 \GeV \lesssim m_{H_0}
\lesssim 80 \GeV$}), and the high mass ($500 \GeV \lesssim m_{H_0}
\lesssim 15 \TeV$) ranges.

Candidates in the low mass range annihilate only through the Higgs
channel. These solutions may be compatible with the DAMA
result~\cite{Andreas:2008xy}, an hypothesis that may be tested in
the future by other direct detection experiments (CDMS, XENON and
CoGeNT). Candidates in the middle mass range annihilate in the $Z$
(when coannihilation with $A_0$ is kinematically allowed) or the
Higgs channel. Some solutions give a very significant gamma ray line
from annihilation at the GC which might be observable with the
GLAST/Fermi satellite~\cite{Gustafsson:2007pc}. Above $80 \GeV$,
annihilation into $W^\pm$ pairs is allowed, the cross section is
large and the abundance falls well below the WMAP abundance
\cite{LopezHonorez:2006gr}. At higher masses, the annihilation cross
section tends to decrease and the relic abundance increases. This
regime is analogous to that of Minimal Dark
Matter~\cite{Cirelli:2005uq} but, in the IDM case, there are more
parameters to play with and there is a whole mass range around
$1$~TeV of candidates with an abundance consistent with
WMAP~\cite{LopezHonorez:2006gr}.

\section{Dark matter distribution} \label{sec-abundance}

The distribution of dark matter in the Galaxy is uncertain, in
particular in the central region. Rotation curve observations
suggest a rather cored profile~\cite{Flores:1994gz,Kravtsov:1997dp},
with a flat behaviour at the centre, whereas numerical simulations
predict more cuspy profiles in the innermost region of the Galactic
centre (see Kravtov et al.~\cite{Kravtsov:1997dp},
Navarro-Frenk-White (NFW)~\cite{Navarro:1995iw} and Moore et
al.~\cite{Moore:1999nt} as well as the recent Via Lactea and
AQUARIUS simulations~\cite{Diemand:2008in,Springel:2008cc}). Other
studies even indicate the existence of a dark
disk~\cite{Read:2008fh,Bruch:2008rx}. In the present study, we will
work with a fixed astrophysical framework and focus on the NFW
profile which is parameterized as follows:
\begin{equation}
\rho(r) = \rho_0 ~\left(\frac{r}{r_0}\right)^{-\gamma} \left[
\frac{1+\left(r_0/a_0\right)^{\alpha}}
{1+\left(r/a_0\right)^{\alpha}}\right]^{\left(
\frac{\beta-\gamma}{\alpha}\right)},
\end{equation}
where $r_0$ is the  distance to the centre of the Galaxy, $\rho_0 =
0.3 \GeV/\xcubcm$ is the dark matter density in the solar
neighbourhood and ($r_0 \ [\xkpc]$, $a_0 \ [\xkpc]$, $\alpha$,
$\beta$, $\gamma$) = (8.5, 20, 1, 3, 1). In this framework, the dark
matter halo is spherical and at rest in the Galactic coordinate
system while the Sun moves on an orbit with mean velocity $v_\odot
\approx 220 \kmpers$. The velocity of dark matter is supposed to
have a Maxwell-Boltzmann distribution with mean velocity $\bar v =
270 \kmpers$.

\section{Indirect detection from the Sun and the Earth} \label{sec-Capture}

\subsection{Capture rate}

Dark matter of the Galactic halo might be captured in the Sun or the
Earth due to the energy loss caused by elastic scattering of dark
matter particles on nuclei. The size, density and composition of the
Sun and the Earth are quite different and so is their ability to
capture DM. WIMPs may generically couple with a nucleus through
axial-vector (spin-dependent, SD) or scalar (spin-independent, SI)
interactions~\cite{Goodman:1984dc}. In the latter case, elastic
scattering with a nucleus is coherent and the cross section is
proportional to the square of its atomic number $A^2$, while
spin-dependent interactions depend only on the total spin of the
nucleus $\sigma \propto J(J+1)$. The Sun is big and mainly composed
of hydrogen atoms ($\sim 77 \%$ of the mass of the Sun). Hydrogen is
the lightest nucleus and therefore capture is likely to be more
important for WIMPs with spin-dependent interactions. On the
contrary, the Earth is comparatively small and mainly composed of
heavy nuclei. Capture of dark matter particles may therefore be
achieved by scalar spin-independent interactions.

Our treatment of capture in the Sun and the Earth will be
elementary. Both for the sake of exploration and because, for the
Earth, we will focus on a dark matter candidate in a mass range
close to the iron mass, we follow the simple approximation
of~\cite{Press:1985ug} and presented in the classical
review~\cite{Jungman:1995df}. This approach is expected to work for
capture in the Sun. However, the situation for the Earth is more
complex. The crux of the problem is that the Earth escape velocity
is relatively small compared to the halo mean velocity, and in
general capture in the Earth is sensitive to the (not well
understood) low velocity part of the dark matter velocity
distribution. The approximation of~\cite{Press:1985ug}
and~\cite{Jungman:1995df} consists in treating capture of dark
matter from the halo as if the Earth was in free space. However, the
Earth is actually deep in the gravitational field of the Sun.
Moreover, either through direct collisions or gravitational
interactions, dark matter from the halo might be captured on solar
bound orbits, possibly affecting the abundance and velocity
distribution of dark matter particles near the Earth position. These
and related issues have been investigated in details by Gould
in~\cite{Gould:1987ww} and further in~\cite{Gould:1999je}, by Damour
and Krauss~\cite{Damour:1998rh}, by Edsj\"{o} and
Lundberg~\cite{Lundberg:2004dn} and, more recently, by Peter
\cite{Peter:2009mi,Peter:2009mm}. To make short a long, albeit
fascinating story\footnote{An historical summary, may be found in
the article by Edsj\"{o} and Lundberg~\cite{Lundberg:2004dn}.} --
which implies taking into account the influence of planets -- let us
summarize here that these works show that the approximation we use
for capture in the Earth works reasonably well if the dark matter
candidate's mass is close to the iron resonance (see figure~16
of~\cite{Lundberg:2004dn} and figure~10 of~\cite{Peter:2009mm}),
which is precisely the case we  focus on in the present work. A more
precise approach is clearly possible, for instance using the
DarkSusy numerical package~\cite{Gondolo:2004sc}, but in the present
work we have limited ourself to an analytical treatment both for
capture in the Sun and in the Earth, an approach which, although not
strictly speaking correct, is very simple to implement.

In the plain Inert Doublet Model, elastic scattering of $H_0$ with
nuclei only takes place through Higgs exchange, which is a scalar,
spin-independent interaction, figure 1.\footnote{For candidates in
the middle and low mass ranges, inelastic scattering with $Z$ boson
exchange ({\em i.e.} $H_0 \rightarrow A_0$) is severely constrained
by direct detection limits which impose $m_{A_0}-m_{H_0} \gtrsim
100$ keV~\cite{Barbieri:2006dq}.} The low energy cross section for
spin-independent scattering of $H_0$ on a nucleus $\mathcal{N}_i$ of
mass $m_{\mathcal{N}_i}$ is
\begin{equation}
\sigma_{H_0\mathcal{N}_i\rightarrow H_0\mathcal{N}_i}^{SI} ~ = ~
\frac{1}{\pi} ~ \frac{\lambda_L^2}{m_h^4} ~
\frac{m_{\mathcal{N}_i}^4}{(m_{H_0} +  m_{\mathcal{N}_i})^2} ~ f^2
\, . \label{eq-sigmaH0NSI}
\end{equation}
The factor $f$, that parametrizes the Higgs to nucleons coupling, is
related to the trace anomaly, $f m_{\mathcal{N}}\equiv \langle
\mathcal{N}| \sum_q m_q
\bar{q}q|\mathcal{N}\rangle=g_{h\mathcal{NN}} v$. This factor is not
well known due to the uncertainty on the contribution of strange
quarks to the nucleon's mass. From the results quoted in
ref.~\cite{Andreas:2008xy}, we take $f=0.30$ as central value for a
nucleon (we do not make a distinction between the coupling of the
Higgs to a proton and a neutron) but it may be within a rather wide
range $0.14  < f <  0.66$.

\begin{figure}[htb!]
\begin{center}
\includegraphics[clip = true, viewport = 8.8cm 23.1cm 12.2cm 26.6cm, width=2.6cm]{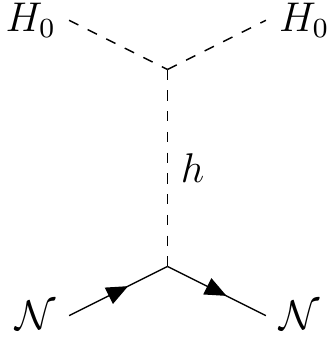}
\vspace{0.3cm} \caption[Feynman diagram for scattering of $H_0$ on a
nucleus $\mathcal{N}$ through exchange of a Higgs boson
$h$]{Scattering of $H_0$ on a nucleus $\mathcal{N}$ through exchange
of a Higgs boson $h$.} \label{fey-H0N-h-H0N}
\end{center}
\end{figure}

The rate at which dark matter is captured by spin-independent
scattering is given by,
\begin{equation} \label{eq-CaptureRate}
C^{\odot/\oplus} \; = ~ c_{\odot/\oplus} ~ \frac{\rho_{local}}{0.3
\GeV \xpercubcm} ~ \frac{270 \kmpers}{\overline{v}_{local}} ~ \sum_i
\, F_i \, f_i \, \phi_i \, S_i \, \frac{\xGeV^2}{m_{\mathcal{N}_i}
\, m_{H_0}} \, \frac{\sigma_i}{10^{-40} \sqcm},
\end{equation}
where $c_\odot ~ = ~ 4.8 \, * \, 10^{24} \pers$ ($c_\oplus ~ = ~ 4.8
\, * \, 10^{15} \pers $) sets the scale for WIMP capture by the Sun
(Earth), ref.~\cite{Jungman:1995df}.

The two astrophysical parameters, $\rho_{local} = 0.3 \GeV$ and
$\overline{v}_{local}=270 \kmpers$, refer to the local density and
mean velocity of the dark matter halo. The sum runs over all
elements present in the respective astrophysical object and encodes
the information regarding their composition. The parameters $f_i$
and $\phi_i$ refer to the mass fraction and the distribution of
element $i$. For those, we use the values given
in~\cite{Jungman:1995df}.

The cross section for capture $\sigma_i = \sigma_i(m_{H_0},m_{N_i})$
via spin-independent scattering on nuclei at low momentum transfer
is given by eq.~(\ref{eq-sigmaH0NSI}). The $F_i = F_i(m_{H_0})$ are
nuclear form factors and the suppression factor $S_i =
S_i\big(\frac{m_{H_0}}{m_{N_i}}\big)$ encodes the kinematics of
DM-nuclei elastic collisions. Capture in the Earth is greatly
enhanced by the appearance of resonances between the masses of the
nuclei and of the WIMP. For the discussion, it is useful to
introduce the dimensionless parameter $x =
m_{H_0}/m_{\mathcal{N}_i}$. The effect of resonances can be
understood from an interplay of two factors: the kinematic
suppression $S_i$ and the form-factor suppression $F_i$. The
kinematic suppression factor $S$ is defined as:
\begin{equation}
S(x) ~ = ~ \Big[ \, \frac{A^{3/2}}{1 \, + \, A^{3/2}} \,
\Big]^{2/3}, \qquad \mathrm{with} \qquad A(x) ~ = ~ \frac{3}{2} \,
\frac{x}{(\, x \, -\, 1)^2} \,\Big( \, \frac{\langle\, v_{esc}^2 \,
 \rangle}{\overline{v}_{local}^2} \, \Big), \label{eq-S}
\end{equation}
where $\langle\, v_{esc}^{\odot} \rangle = 1156 \unit{km/s}$
($\langle\, v_{esc}^{\oplus} \rangle = 13.2 \unit{km/s}$) is the Sun
(Earth) escape velocity.

The behaviour of $S$ depends on the escape velocity and therefore
differs for Sun and Earth. $S$ asymptotically scales like $A$ when
$A \rightarrow 0$ while $S \rightarrow 1$ for $A \rightarrow
\infty$. $A$ is strongly peaked when the WIMP's mass matches a
nucleus' mass ($x \rightarrow 1$). For the Sun, the higher escape
velocity brings $A$ above 1 in all but a few extreme points (small
$m_{H_0}$) for all elements except hydrogen. Because of this
behaviour, $S \simeq 1$ in the biggest part of the mass range.
However, in the case of the Earth, $A \ll 1$ almost everywhere
except at the resonances. Therefore, $S \simeq 1$ only in a small
region around the resonance while elsewhere $S \sim A \ll 1$.

The nuclear suppression form factor takes into account the fact that
nuclei are made of nucleons. If both the WIMP's kinetic energy in
the halo and its maximum energy transfer to the nucleus are
comparable to the characteristic coherence energy, the WIMP sees the
substructure of the nucleus and the cross section is reduced. For
the Earth, it can be derived~\cite{Gould:1987ir} that this effect is
only relevant at the resonances and that the suppression is
negligible ($5\%$ effect) for all elements except iron. Therefore,
following~\cite{Jungman:1995df}, we assume that the form factor
suppression is 1 for all elements but iron for which it is
parameterized as
\begin{equation}
F_{\mathrm{Fe}} ~ = ~ 1 \, - \, 0.26 \frac{A}{1 + A} \, .
\label{eq-FIronEarth}
\end{equation}

In the case of capture by the Sun, one has to take into account that
the kinetic energy of the WIMP when scattering on a nucleus is not
anymore the kinetic energy at infinity (connected with the local
velocity in the halo) but receives a contribution from the Sun's
gravitational potential. It can then be shown that the form-factor
suppression is no longer confined to the resonances and additionally
that it applies to all elements~\cite{Jungman:1995df}:
\begin{equation}
F_i ~ = ~ F_i^{inf} \, + \, (1 - F_i^{inf}) \, \exp \Big[ - \Big(
\frac{\log m_{H_0}}{\log m_c^i} \Big)^{\alpha_i} \Big]
\label{eq-FSun}
\end{equation}
except for hydrogen, where $F = 1$. Therein $F_i^{inf}$, $\m_c^i$
and $\alpha_i$ are fit parameters tabulated
in~\cite{Jungman:1995df}. The form-factor in the Sun has basically
no effect on the lightest elements (hydrogen and helium) while for
heavier nuclei it leads to a suppression which is increasing with
the mass of the element. Capture on iron experiences the strongest
suppression, ${\cal O}(10^{-1}-10^{-2})$ for the heaviest dark
matter candidates in the middle range under consideration.

\begin{figure}[htb!]
\begin{center}
\includegraphics[clip = true, viewport = 5.0cm 17.2cm 15.8cm 25.6cm, width=9cm]{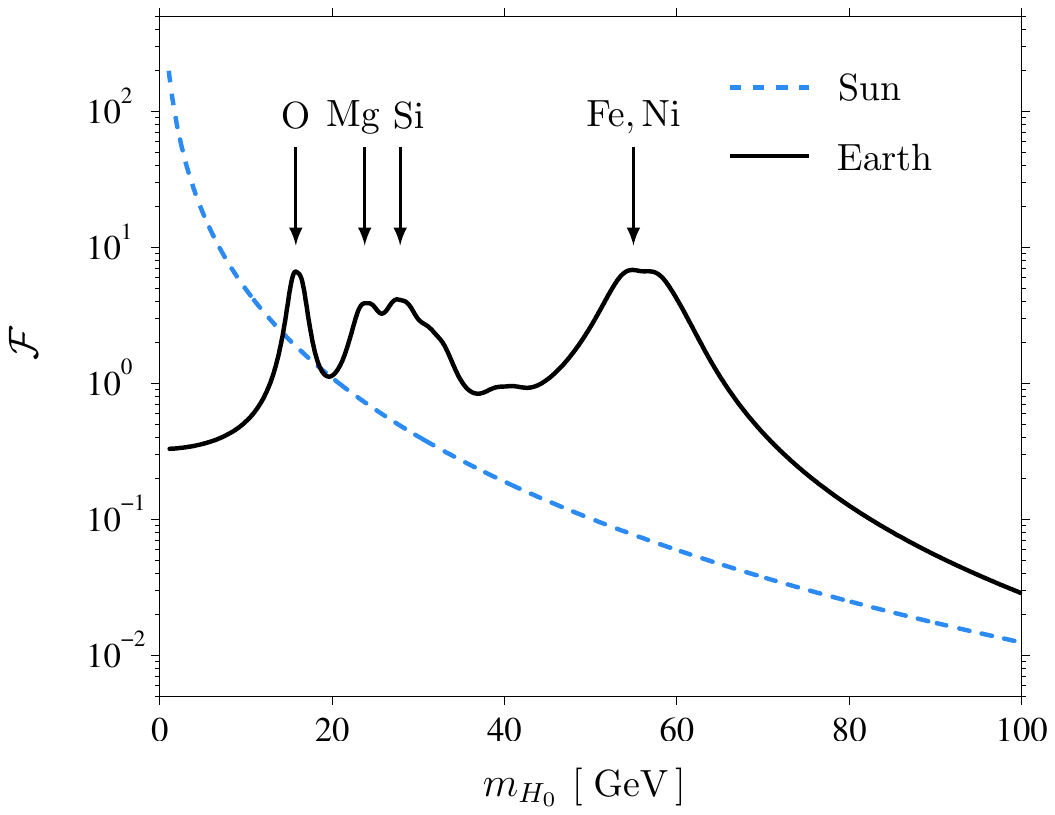}
\caption[${\cal F}$ in function of $m_{H_0}$ for Sun and
Earth]{Typical behaviour of ${\cal F}_{\odot/\oplus}$ in function of
$m_{H_0}$ for Sun (dashed blue line) and Earth (solid black line).
The resonance phenomenon, which is present for elements in the
Earth, is absent for the Sun (here for $m_h=120 \GeV$, $f=0.3$ and
$\lambda_L=0.2$). \label{fig-f}}
\end{center}
\end{figure}

To summarize the dependence on the WIMP's mass and the different
nuclei, we define for the IDM candidate a function
$\mathcal{F}_{\odot/\oplus}$
\begin{equation}
\mathcal{F}_{\odot/\oplus} =  \sum_i f_i^{\odot/\oplus}
\phi_i^{\odot/\oplus}
S\big(\frac{m_{H_0}}{m_{\mathcal{N}_i}},v_{esc}^{\odot/\oplus}\big)
\, F_i^{\odot/\oplus}(m_{H_0},\mathcal{N}_i) \,
\frac{m_{\mathcal{N}_i}^3}{( m_{H_0} + m_{\mathcal{N}_i} )^2} \,
\frac{\lambda_L^2}{m_{H_0} m_h^4} \, f^2 \ \frac{\xGeV^2}{10^{-40}
\sqcm}. \label{eq-f}
\end{equation}
The behaviour of $\mathcal{F}_{\odot/\oplus}$ as a function of
$m_{H_0}$ is displayed in figure~\ref{fig-f} both for the Earth and
the Sun. It shows in particular the great importance of the iron
resonance in the case of the Earth. With the definition of
$\mathcal{F}_{\odot/\oplus}$, the capture rate can be written:
\begin{equation}
C^{\odot/\oplus} ~~ = ~~ \frac{c_{\odot/\oplus}}{\pi} ~~
\frac{\rho_{local}}{0.3 \GeV \percubcm} ~~ \frac{270
\kmpers}{\overline{v}_{local}} ~~ \mathcal{F}_{\odot/\oplus}.
\label{eq-CaptureRatef}
\end{equation}

\subsection{Annihilation rate}

\begin{figure}[htb!]
\begin{center}
\includegraphics[clip = true, viewport = 8cm 17.6cm 13cm 20.2cm, width=3.9cm]{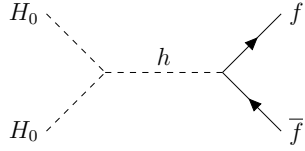}
\vspace{0.3cm} \caption[Feynman diagram for annihilation of $H_0$
into fermions.]{The only tree level annihilation channel of light
$H_0$ ($m_{H_0} \lesssim m_W$) into fermions pairs is through the
Higgs boson.}\label{fey-H0H0-h-ffbar}
\end{center}
\end{figure}

The number $N$ of WIMP dark matter in the Sun or Earth is controlled
by the capture and annihilation rates
\begin{equation}
\frac{dN}{dt} ~ = ~ C ~ - ~ 2 \, \Gamma, \label{eq-dNdtSE}
\end{equation}
where $C$ is the capture rate defined in eq.~(\ref{eq-CaptureRate})
and $\Gamma$ is the annihilation rate. For a self conjugate
particle, like $H_0$, the total annihilation rate from a (effective)
volume $V_{\mathrm{eff}}$ is given by
\begin{equation}
\Gamma ~ = ~ \frac{N^2}{2 V_{\mathrm{eff}}} ~ \sigma v.
\label{eq-GammaSEVeff}
\end{equation}
The cross section for annihilation of $H_0$ is through the Higgs
channel only, figure~\ref{fey-H0H0-h-ffbar}, and is given by
\begin{equation}
\sigma |\vec{v}|_{H_0H_0\rightarrow f\bar{f}}  ~ = ~ \frac{n_c}{\pi}
~ \frac{\lambda_L^2}{( \, 4 \, m_{H_0}^2 \, - \, m_h^2 \, )^2} ~
\frac{m_f^2 \, ( \, m_{H_0}^2 \, - \, m_f^2 \,
)^{3/2}}{m_{H_0}^3}\,. \label{eq-sigmaH0f}
\end{equation}
With eq.~(\ref{eq-GammaSEVeff}) the solution of~(\ref{eq-dNdtSE}) is
\begin{equation}
\Gamma ~ = ~ \frac{C}{2} ~ \tanh^2\Big(\, \sqrt{\frac{\sigma
v}{V_{\mathrm{eff}}} ~ C} ~ t_{\odot} \, \Big) ~ = ~ \frac{C}{2} ~
\tanh^2\big(\, t_{\odot} \, / \, \tau \, \big) ~ = ~ \frac{C}{2} ~
F_{\mathrm{EQ}}, \label{eq-GammaSEMaj}
\end{equation}
where
\begin{equation}
F_{\mathrm{EQ}} ~ = ~ \tanh^2\big(\, t_{\odot} \, / \, \tau \, \big)
\qquad \mathrm{and} \qquad \tau ~ = ~
\sqrt{\frac{V_{\mathrm{eff}}}{\sigma v \; C}}, \label{eq-FEQ}
\end{equation}
and $t_{\odot} = 4.5 \Gyr$ is the age of the solar system. The
effective volume of the Sun or of the Earth is determined by roughly
matching the core temperature with the gravitational potential
energy of the WIMP at the core radius.
Following~\cite{Jungman:1995df,Gould:1987ir,Halzen:2005ar} we adopt
\begin{eqnarray}
V_{\mathrm{eff}}^{\odot} ~ & = & ~ 5.8 * 10^{30} \, \cubcm ~ \Big(
\,
\frac{\GeV}{m_{H_0}} \, \Big)^{3/2}, \label{eq-VeffSun}\\
V_{\mathrm{eff}}^{\oplus} ~ & = & ~ 1.8 * 10^{27} \, \cubcm ~ \Big(
\, \frac{\GeV}{m_{H_0}} \, \Big)^{3/2}. \label{eq-VeffEarth}
\end{eqnarray}

If the time scale $\tau$ for capture and annihilation to equilibrate
is small compared to the age of the solar system, the equilibrium
factor and annihilation rate reach a maximum, $t_{\odot} \, \gg \,
\tau \; \Rightarrow \; F_{\mathrm{EQ}} \, = \, 1 \; \Rightarrow \;
\Gamma \, = \, C / 2$.

Given $\Gamma$, the differential flux of annihilation products $i$
in the Sun or Earth is given by
\begin{eqnarray}
\Big(\, \frac{d\phi}{dE} \, \Big)_i & = & \frac{\Gamma}{4 \pi R^2} ~
\sum_F ~ BR_F ~ \Big(\, \frac{dN}{dE} \, \Big)_{F,i}
\label{eq-NuFluxSE-Gamma} \\
& = & \frac{1}{2} ~ \frac{C \, F_{\mathrm{EQ}}}{4 \pi R^2} ~ \sum_F
~ BR_F ~ \Big(\, \frac{dN}{dE} \, \Big)_{F,i},
\label{eq-NuFluxSE-CFEQ}
\end{eqnarray}
where $R$ is either the distance from the Sun to the Earth ($R \, =
\, d_{\odot-\oplus} \, = \, 1.5 * 10^8 \km$) or the radius of the
Earth ($R \, = \, r_{\oplus} \, = \, 6300 \km$). $BR_F$ is the
branching ratio for annihilation into final state $F$ and
$(dN/dE)_{F,i}$ is the differential spectrum of the annihilation
products $i$ through the decay channel $F$. The sum on $F$ runs over
all the kinematically allowed final states.

\subsection{Muon flux} \label{sec-MuFluxSE}

In the following, only the resulting flux of muonic neutrinos will
be taken into account, since they provide the best directional
information. Electron- or tau-neutrinos produce a particle shower
close or within the detector volume hence blurring the directional
information.

For neutrinos from dark matter annihilation, one may distinguish two
extreme cases referred to as soft and hard spectra. The soft
spectrum is falling rapidly with the neutrino energy. The hard
spectrum on contrary is flatter with significant contributions at
high energies. Specifically, annihilations of $H_0$ into  $b
\overline{b}$ quarks give a soft spectrum of neutrinos, while a hard
spectrum originates either from pairs of $W$ or $Z$ bosons (if
$m_{H_0} \gtrsim m_W$) or from $\tau^+ \tau^-$ (if $m_{H_0} < m_W$)
channels.

The detection rate per unit detector area of neutrino-induced,
through-going muon events from the Sun or the Earth is computed
using the charged-current cross sections and the muon range.
Ignoring detector thresholds, as in reference~\cite{Jungman:1995df},
the muon flux in the detector is found to be:
\begin{equation} \label{eq-MuFluxSE}
\phi_\mu^{\odot/\oplus} ~ = ~ \, \varphi_{\odot/\oplus} ~~
\frac{C^{\odot/\oplus} \, F_{\mathrm{EQ}}}{2 \pers} ~ \Big( \,
\frac{m_{H_0}}{\GeV} \, \Big)^2 ~ \sum_i \, a_i \, b_i ~ \sum_F \,
BR_F \, \langle \, N z^2 \rangle_{F,i} \, (m_{H_0}),
\end{equation}
where $\varphi_\odot ~ = ~ 2.54 \, * \, 10^{-23} \persqkm \peryr$
($\varphi_\oplus ~ = ~ \varphi_\odot ~ ( d_{\odot-\oplus} \, /
r_{\oplus} )^2  ~ = ~ \varphi_\odot ~ * ~ 5.6 \, * \, 10^{6}$) for
the Sun (Earth), $C^{\odot/\oplus}$ is the capture rate defined in
eq.~(\ref{eq-CaptureRate}, \ref{eq-CaptureRatef}) and
$F_{\mathrm{EQ}}$ is the equilibrium factor defined in
eq.~(\ref{eq-FEQ}). The neutrino-scattering coefficients $a_i$ and
muon-range coefficients $b_i$ are given in~\cite{Jungman:1995df} as
$a_\nu \, = \, 6.8$, $a_{\overline{\nu}} \, = \, 3.1$, $b_\nu \, =
\, 0.51$ and $b_{\overline{\nu}} \, = \, 0.67$. $BR_F$ is the
branching ratio for annihilation into final state $F$ and $\langle
\, N z^2 \rangle_{F,i}(m_{H_0})$ is the second moment of the
spectrum $(dN/dE)_{F,i}$ of neutrinos of type $i$ from final state
$F$ scaled by the square of the injection energy $E_{in}$
\begin{equation}
\langle \, N \, z^2 \, \rangle_{F,i} \,(E_{in}) ~ = ~
\frac{1}{E_{in}^2} \, \int \, \Big( \, \frac{dN}{dE} \, \Big)_{F,i}
\, (E_{\nu}, E_{in}) \, E_{\nu}^2 \, dE_{\nu}. \label{eq-IntNuSpec}
\end{equation}
As neutrino detectors have certain thresholds on the minimum
neutrino energy detectable, a lower, detector dependent, energy
bound is implicit in the integration range.

\section{Indirect detection from the Galactic centre} \label{sec-NuFluxGC}

Simulations indicate that there is a higher dark matter density at
the centre of the Milky Way. As for the Sun and the Earth, this
opens the possibility of indirect detection of dark matter through
its annihilation products. The expected flux from the Galactic
centre depends on the assumed dark matter density profile $\rho$.

For a self conjugate dark matter candidate, like the $H_0$, the flux
of annihilation products $i$ at the Earth is
\begin{eqnarray}
\frac{d\phi_i}{dE} ~ &=& ~ \frac{1}{2} ~ r_0 \, \rho_{local}^2 ~
\frac{dN_i}{dE} ~ ~ \frac{\sigma v}{\, 4 \pi m_{H_0}^2} ~ ~
\overline{J}(\Delta\Omega)
~ \Delta\Omega \label{eq-diffFluxGC},\\
\mathrm{with} ~~~~ \overline{J}(\Delta\Omega) & = &
\frac{1}{\Delta\Omega} ~ \frac{1}{r_0 \, \rho_{local}^2} ~
\int_{\Delta\Omega} d\Omega ~~ \int_{\mathrm{line \ of \ sight}} ds
~~ \rho^2(r(s,\theta)), \label{eq-Jbar}
\end{eqnarray}
where $r_0 = 8.5 \kpc$ is the distance to the Galactic centre,
$dN/dE$ is the energy spectrum of annihilation products and $\sigma
v$ is the total annihilation cross section times velocity. The
factor $1/2$ results from the number of ways to have a pair out of
$n$ particles ${\sim}\, n^2/2$.\footnote{This factor of $1/2$ is in
agreement with the expression given by~\cite{Taoso:2007qk},
\cite{Ullio:2002pj} and~\cite{cirelli-2008}.} The quantity
$\overline{J}$ encompasses all the astrophysics and is defined as
the integral of $\rho^2(r)$ along the line of sight averaged over a
solid angle $\Delta\Omega$. With a NFW profile, the integration
gives for instance $\overline{J} = 1382$ for $\Delta \Omega =
10^{-3}$, corresponding to the resolution of EGRET, or $\overline{J}
= 536$ for $\theta = 2.5^{\circ}$, which is the resolution of
ANTARES, a neutrino telescope which is able to look in the direction
of the Galactic centre.

\section{IDM results} \label{sec-IDMresults}
Our discussion is divided into the three mass ranges for the $H_0$
candidate. In each case, the relevant astrophysical context is
distinct.

\subsection{Low mass candidate: indirect detection from the Sun}

The light dark matter candidate $m_{H_0} < 10 \GeV$ of the IDM has
been studied in~\cite{Andreas:2008xy} in view of explaining the DAMA
annual modulation signal. Such a candidate has a larger number
density than heavier candidates. It is however too light to give a
large signal from the core of the Earth (see figure~\ref{fig-f}).
Because of its large number density and quite large cross section,
annihilations at the Galactic centre might be important and, indeed,
a large flux of photons is to be expected which might be studied by
the GLAST/Fermi satellite~\cite{Andreas:2008xy,Feng:2008dz}. At
these energies, however, the flux of neutrinos is well below the
background of atmospheric neutrinos. The only possible relevant
source of neutrinos might thus be the Sun.

\begin{figure}[htb!]
\begin{center}
\includegraphics[clip = true, viewport = 0.8cm 6.0cm 28.5cm 16.9cm, width=16cm]{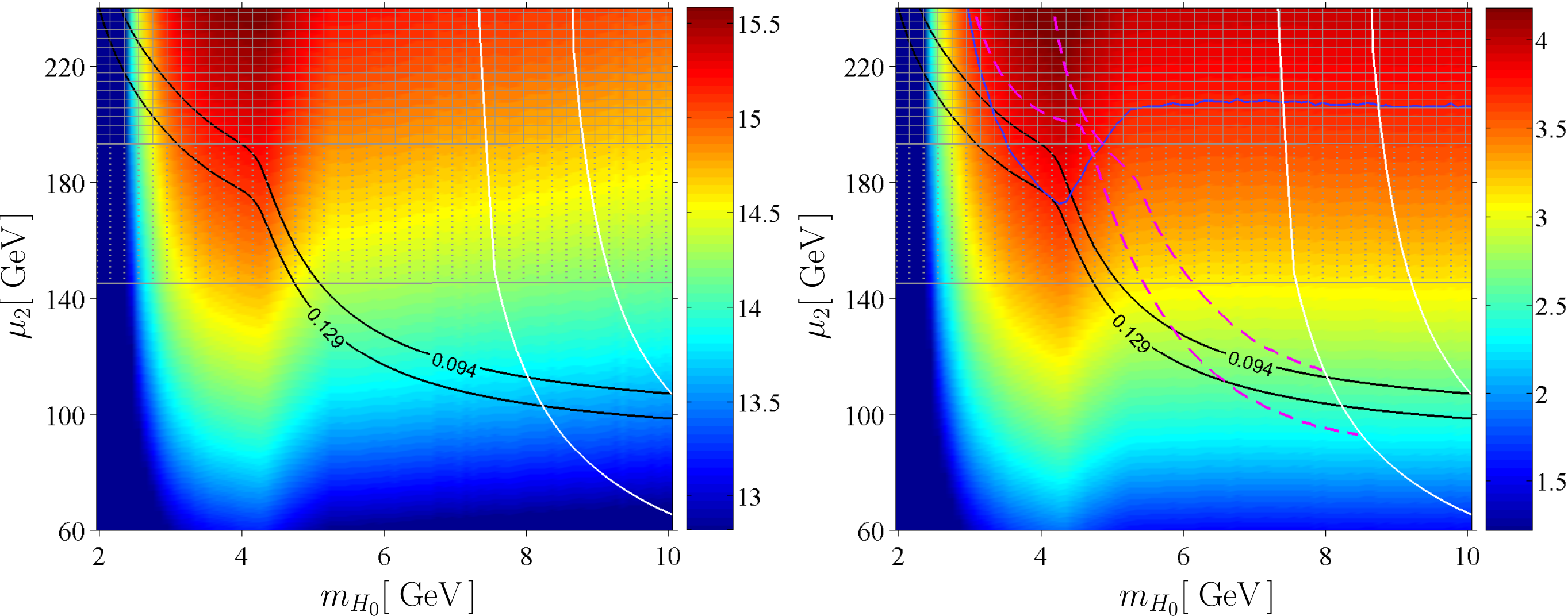}
\caption[Neutrino and Muon Fluxes from the Sun]{Expected neutrino
(\textit{left}) and muon (\textit{right}) fluxes from the Sun, for a
neutrino energy threshold of 2 GeV. Dark blue line in the muon flux
plot: estimate of  Super-Kamiokande sensitivity, conservatively
extrapolated to light WIMPs (see text).\\
\textsl{Colour gradient - $\, \log_{10} \phi_\nu ~ [\textsl{km}^{-2}
\, \textsl{yr}^{-1}]$ \textit{(left)} and $\log_{10} \phi_\mu ~
[\textsl{km}^{-2} \, \textsl{yr}^{-1}]$ \textit{(right)}; WMAP area
(black lines); DAMA allowed region (dashed magenta lines); XENON,
CDMS exclusion limits (white lines, from left to right); The
shaded regions correspond to $\lambda_2 > 1$. \\
(Parameters: $m_h = 120 \, \textsl{GeV}$, $\lambda_2 = 1$ (dotted)
and $\lambda_2 = \pi$ (shaded),  $f = 0.3$).}
\label{fig-NuMuFluxSSuperK}}
\end{center}
\end{figure}

The result of our analysis, for $m_h = 120 \GeV$, is given in
figure~\ref{fig-NuMuFluxSSuperK} where we show both the expected
flux of muonic neutrinos according to eq.~(\ref{eq-NuFluxSE-CFEQ})
and that of muons from eq.~(\ref{eq-MuFluxSE}) as a function of the
mass of the dark matter candidate, $m_{H_0}$, and the bare mass
scale, $\mu_2$. The neutrino spectra have been computed using
\textsc{WimpSim}~\cite{Wimpsim,Blennow:2007tw}\footnote{To take into
account the effect of neutrino oscillations, we take the flux of
muon neutrinos at the Earth to be equal to one third of the total
flux of neutrinos produced at the core of the Sun.} and are cut at a
threshold energy $E_\nu = 2 \GeV$. The fluxes are shown together
with the WMAP allowed region in black (which we have computed using
\textsc{micrOMEGAs}~\cite{Belanger:2006is}) for $0.094 < \Omega_{DM}
h^2 < 0.129$, the current limits set by direct detection experiments
(in white, from left to right, XENON and CDMS respectively) and the
DAMA allowed region (in light blue) for $f = 0.3$ (cf.
ref.~\cite{Andreas:2008xy}, the allowed region remaining after
exclusions of all other direct searches).

As emphasized in~\cite{Feng:2008qn} and~\cite{Hooper:2008cf}, the
flux of neutrinos produced in the Sun by such a light DM candidate
might be constrained by the Super-Kamiokande experiment. Its
sensitivity is currently only set down to $m_{DM} = 18 \GeV$ from
studies of through-going muons~\cite{Desai:2004pq}. It could however
be extended down to 2 GeV including stopping, partially contained or
fully contained muons~\cite{Feng:2008qn}. In the present work, the
estimation of the Super-Kamiokande sensitivity (drawn in blue) has
been obtained horizontally extrapolating the one given
in~\cite{Desai:2004pq} to lower DM masses. This limit has been
rescaled to a muon threshold of 1 GeV (cf.~\cite{Hubert:2007zza})
which corresponds to the neutrino threshold $E_\nu = 2 \GeV$
assuming an energy loss of $\simeq 50 \ \%$ in the neutrino to muon
conversion. Compared to the projected Super-Kamiokande sensitivity
presented in~\cite{Feng:2008qn}, our extrapolation is more
conservative.

Since $m_{H0}^2 = \mu_2^2 + \lambda_L v^2$, a light $H_0$ and
$\mu_2$ in the range of both DAMA and WMAP require a large, negative
coupling $\lambda_L$ within $\sim [ -0.6, -0.1]$. This in turn
requires a rather large $\lambda_2$ to insure the stability of the
potential. Concretely one needs~\cite{Barbieri:2006dq}
\begin{equation}
\lambda_2 \ > \ \lambda_L^2 \ / \ \lambda_1 \ \equiv \ \frac{2
(m_{H_0}^2 - \mu_2^2)^2}{v^2 m_h^2}. \label{eq-lambda2}
\end{equation}
for $\lambda_{1,2} >0$ and $\lambda_L <0$. In
figure~\ref{fig-NuMuFluxSSuperK}, the shaded regions correspond to a
coupling $\lambda_2 > 1$, for $m_h = 120 \GeV$. In the low mass
range, both the annihilation and the capture cross section are
proportional to $\lambda_L^2/m_h^4 \propto \mu_2^4/m_h^4$
($\lambda_L \propto \mu_2^2$), cf. equations~(\ref{eq-sigmaH0NSI})
and~(\ref{eq-sigmaH0f}). When increasing the Higgs mass, this ratio
must be kept constant not to reduce the neutrino/muon flux. The
Super-Kamiokande sensitivity is therefore shifted to higher values
of $\mu_2$ that are increased $\propto m_h$. On the other hand, the
displacement of the exclusion limit $\lambda_2 > 1$ is only $\propto
\sqrt{m_h}$. The sensitivity lying for $m_h = 120 \GeV$ in the
$\lambda_2 > 1$ region will require even larger values of
$\lambda_2$ for higher Higgs masses.

Figure~\ref{fig-NuMuFluxSSuperK} shows that Super-Kamiokande is
potentially able to test the light IDM candidate and a part of the
DAMA allowed region (from about 3 GeV to 4.8 GeV). However, the
required couplings tend to be rather large.

\begin{figure}[htb!]
\begin{center}
\includegraphics[clip = true, viewport = 2.1cm 8.8cm 12.4cm 16.8cm, width=9cm]{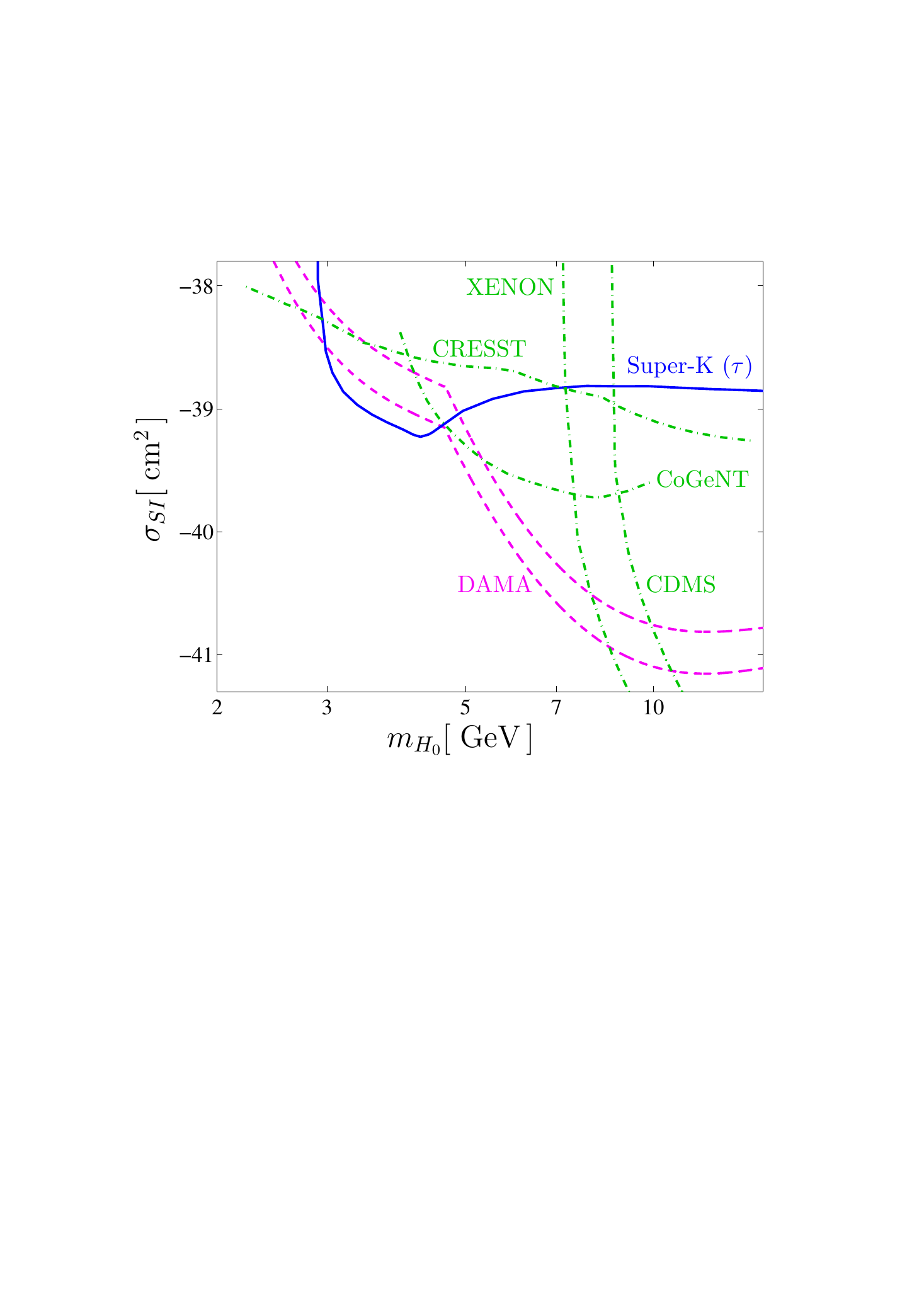}
\end{center}
\caption{The region allowed by DAMA (dashed magenta lines) together
with limits on the spin-independent elastic scattering cross section
of $H_0$ on a nucleus from direct detection experiments (dash-dotted
green lines), cf.~\cite{Petriello:2008jj}. Super-Kamiokande (solid
blue line) sets another limit from the $\tau^+\tau^-$ annihilation
channel of $H_0$. This limit results from the one presented
in~\cite{Hooper:2008cf} rescaled by the IDM branching ratio.}
\label{fig-SuperKtaulimit}
\end{figure}

A similar conclusion is found adapting the analysis of
ref.~\cite{Hooper:2008cf} which gives the limit set by
Super-Kamiokande assuming a DM candidate that annihilates with
branching ratio $=1$ either into $\tau-\overline{\tau}$,
$c-\overline{c}$ or $b-\overline{b}$ pairs, the strongest constraint
being for annihilations into $\tau$ leptons. This limit for
$\tau^+\tau^-$ rescaled by the actual branching ratio in the frame
of the IDM is shown in figure~\ref{fig-SuperKtaulimit} (in blue)
together with the DAMA allowed region (in light blue) and other
direct detection experiments (in green).

For the $H_0$ candidate, the limit from the tau channel indicates
(in agreement with the result of figure~\ref{fig-NuMuFluxSSuperK})
that the limit set by Super-Kamiokande is competitive with limits
set by direct detection experiments (modulo the uncertainties
regarding the abundance of dark matter in the Sun versus that at the
Earth, as usual).

\subsection{Middle mass candidate: indirect detection from the Earth}
\label{sec-middle}

This refers to a $H_0$ with a mass typically between 40 and 80
GeV~\cite{LopezHonorez:2006gr}. Since the $H_0$ has essentially
spin-independent interactions, the best place to look for neutrinos
produced in annihilations is toward the centre of the Earth.
However, direct detection experiments put very stringent limits on
the SI cross section of WIMPs and so on the potential signatures in
neutrino telescopes.

In this section, we discuss the muon flux resulting from $H_0$
annihilations in the Earth. First we consider the Inert Doublet
Model in its most minimalistic version and show that some of the
candidates that have an abundance in agreement with WMAP give a muon
flux that almost reaches the sensitivity of a km size detector like
IceCube. We then consider two extensions. First we discuss the
possible effect of one-loop corrections, in the spirit
of~\cite{Gustafsson:2007pc} but conclude that they may not
significantly improve the signatures of the IDM into neutrinos.
Finally, we consider the extension of the IDM with three heavy
Majorana neutrinos introduced in~\cite{Ma:2006km}. This extension
opens the possibility of annihilations of $H_0$ into two SM $\nu$ or
two $\bar \nu$ with energy $m_{\nu,\bar \nu} = m_{H_0}$, thus
strongly (albeit to the price of some fine tuning) increasing the
potential signature in neutrino telescopes.

\subsubsection{The minimal scenario} \label{sec-IDMbasic}

\begin{figure}[htb!]
\begin{center}
\includegraphics[clip = true, viewport = 2.1cm 15.3cm 17.1cm 27.6cm, width=16cm]{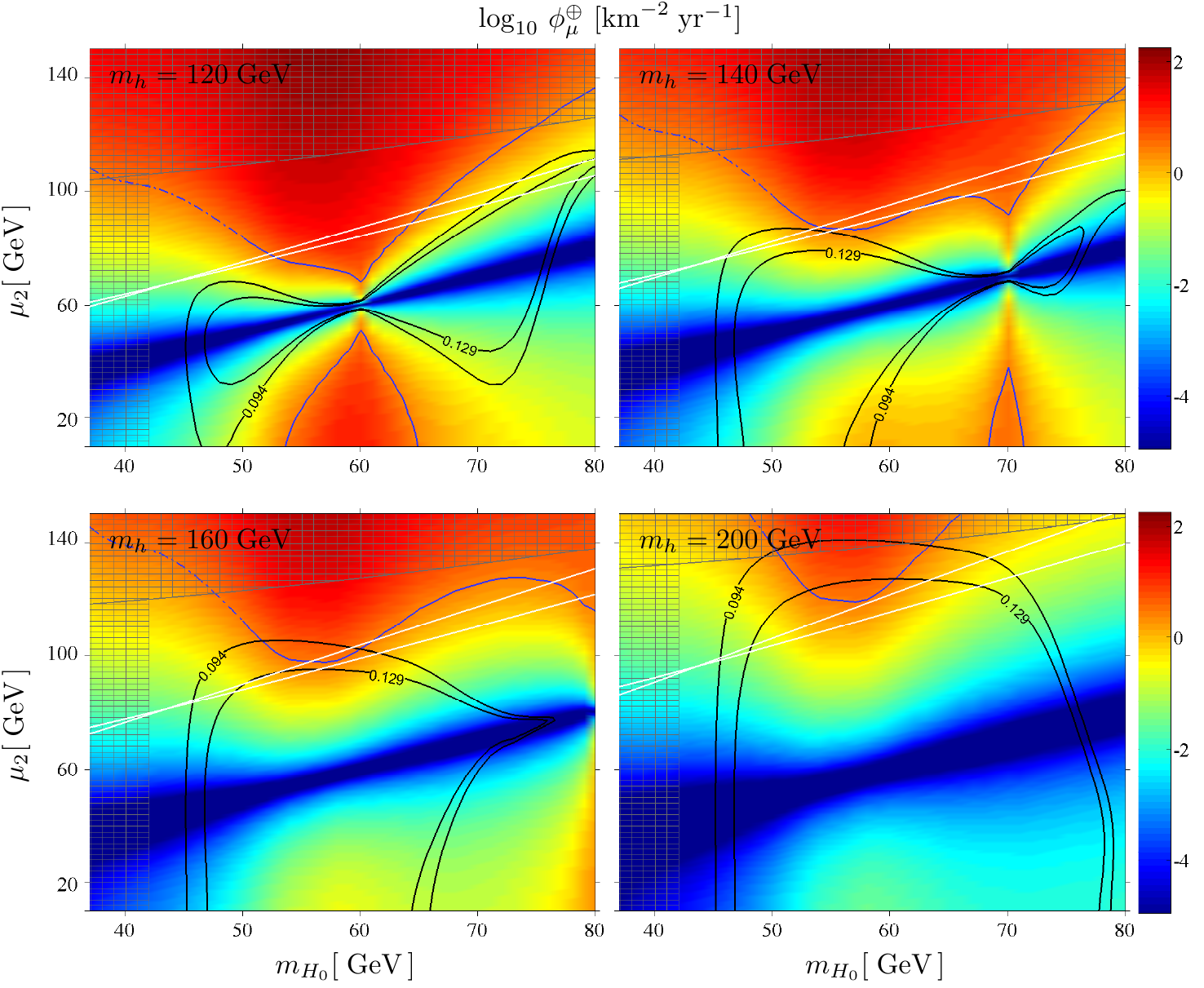}
\caption[Muon Flux from the Earth for the minimal scenario]{Muon
flux from the Earth for the minimal scenario (see text).
No detector threshold is applied. \\
\textsl{Colour gradient - $\, \log_{10} \phi_\mu ~ [\textsl{km}^{-2}
\, \textsl{yr}^{-1}]$; WMAP area (black lines); XENON, CDMS
exclusion limits (white lines); IceCube sensitivity soft
(blue lines); excluded regions (shaded). \\
(Parameters: $\lambda_2 = 0.2$, $\Delta m_{A_0 H_0} = 8 \,
\textsl{GeV}$, $\Delta m_{H^+ H_0} = 50 \, \textsl{GeV}$, $f =
0.3$)}. \label{fig-MuFluxE-bm}}
\end{center}
\end{figure}

Applying the results of section~\ref{sec-MuFluxSE}, the muon flux is
given by~(\ref{eq-GammaSEMaj}). For SI scattering, the Earth is
expected to have a higher capture rate for dark matter masses near
the main resonances. In the range of $H_0$ masses under
consideration, the dominant resonance is that of iron. The expected
muon flux is presented in figure~\ref{fig-MuFluxE-bm} for four
different Higgs masses. In this figure, we neglect the effect of
detector thresholds and give the total flux of muons. Hence, this is
the maximal signal that one may get, of course assuming standard
hypotheses regarding the abundance of dark matter at the centre of
the Earth. The mass splittings for $A_0$ and $H^{\pm}$ (8 and 50 GeV
respectively) are consistent with LEP data~\cite{Lundstrom:2008ai}.

The figures show the muon flux (colour gradient) together with the
$2\sigma$ WMAP abundance region. The shaded regions are excluded by
stability considerations and the LEP I limit on the $Z$ width. We
give the limit (in white) imposed by direct detection experiments
({\em i.e.} CDMS and XENON~\cite{collaboration-2008}). Finally, the
sensitivity of the IceCube detector is shown in blue. It depends on
the neutrino spectrum, for which a distinction between so-called
'hard' and 'soft' spectra is made. The former results from
annihilation channels like $\tau^+ \tau^-$, the latter from $b
\overline{b}$ pairs. For the Earth, the IceCube best case
sensitivity for $m_{DM} = 50 \GeV$ is $\sim 15$ muons per $\sqkm
\yr$ for a hard spectrum and $\sim 40$ muons per $\sqkm \yr$ for a
soft spectrum. Here we adopt the best case sensitivity presented for
example in~\cite{Hubert:2007zza} as a function of $m_{DM}$ for $50
\GeV < m_{H_0} < 100 \GeV$ and its extrapolation down to candidates
with a mass of $30 \GeV$. Since the dominant annihilation channel is
into $b \overline{b}$, we use the sensitivity for a soft spectrum.

Annihilation rates and thus muon fluxes are higher for candidates
with a mass close to $m_h/2$. This however gives a cosmic abundance
too low compared to WMAP. This tension resolves as the mass of the
Higgs is increased. Our results are in agreement with those
of~\cite{Agrawal:2008xz}. However, we find that the candidates with
a potential signature in IceCube correspond to regions of parameters
that are not favoured since they either give a too small abundance
($m_h = 120 \GeV$ and $140 \GeV$), or are within the exclusions
limits set by current direct detection experiments ($m_h = 160 \GeV$
and $200 \GeV$). For the sake of completeness, we illustrate the
variation of the muon flux obtained by varying the Higgs-nucleon
coupling form factor $f$ in figure~\ref{fig-MuFluxE-bm-f}.

\begin{figure}[htb!]
\begin{center}
\includegraphics[clip = true, viewport = 2.1cm 6.2cm 17.1cm 12.6cm, width=16cm]{Figures/Earth_bm.pdf}
\caption[Muon flux from the Earth for the minimal scenario,
different values of the nuclear form factor $f$]{Muon flux from the
Earth for the minimal scenario for different values of the factor
$f$. The figure comes from the down-left plot in
figure~\ref{fig-MuFluxE-bm}, when varying $f$. No detector threshold
is applied.\\
\textsl{Colour gradient - $\, \log_{10} \phi_\mu ~ [\textsl{km}^{-2}
\, \textsl{yr}^{-1}]$; WMAP area (black lines); XENON, CDMS
exclusion limits (white lines); IceCube sensitivity soft
(blue lines); excluded regions (shaded). \\
(Parameters: $\lambda_2 = 0.2$, $\Delta m_{A_0 H_0} = 8 \,
\textsl{GeV}$, $\Delta m_{H^+ H_0} = 50 \, \textsl{GeV}$, $f = 0.14$
(\textit{left}) and $f = 0.66$ (\textit{right}))}.
\label{fig-MuFluxE-bm-f}}
\end{center}
\end{figure}

For small values of $f$, a part of the parameter space with muon
flux above the sensitivity is in agreement with direct detection.
This region is however outside the $2\sigma$ WMAP area. Relaxing the
WMAP limit on the abundance (acceptance obtained with the $3\sigma$
WMAP area) could give a muon flux above the IceCube sensitivity for
some candidates that are not yet excluded by direct detection
experiments.

\subsubsection{A non-minimal scenario: effect of loop corrections} \label{sec-Loop}

To increase the flux of muons one possible strategy is to envision
candidates with a hard(er) neutrino spectrum, thus exploiting the
fact that the muon flux is $\propto E_\nu^2$,
eq.~(\ref{eq-MuFluxSE}). The best case is to have annihilations into
mono-energetic neutrinos, a solution that we will consider in
section~\ref{sec-Majorana}. First we consider the possibility that
$H_0$ may annihilate into $\gamma-\bar \nu \nu$ at one-loop.

At one loop the $H_0$ may annihilate into a pair of photons ($\gamma
\gamma$) or into a photon and a $Z$ boson ($\gamma Z$). The
relevance of these contributions in the IDM model has been
emphasized in~\cite{Gustafsson:2007pc}. In turn, the $Z$ may decay
into $\nu\bar\nu$. There is another contribution, through a loop
with $W$-bosons, for annihilation into $\nu\bar \nu\gamma$. This
contribution has not yet been computed in the literature. Here, to
explore whether it would be worth being computed, we make the
reasonable assumption that this contribution does not overwhelm that
from $Z$ decay.  The energy at which the $Z$ boson is produced is
$E_Z = m_{H_0} + m_Z^2 / 4 m_{H_0}$ if $m_{H_0}
> m_Z/2$ and the spectrum of neutrinos is peaked around $E_Z/2$.

In~\cite{Gustafsson:2007pc}, the possibility of direct annihilation
into photons and $Z$ bosons has been investigated in view of
observing gamma rays from the Galactic centre and four benchmark
models (BM), shown here in table~\ref{tab-BenchmarkBergstrom}, are
proposed.

To compute the expected muon flux for those BMs, all we need is the
decay spectrum of a $Z$ boson in addition to the information given
in table~\ref{tab-BenchmarkBergstrom}. For a $Z$ with an energy
$E_Z$, the neutrino spectrum is obtained using \textsc{WimpSim}. The
muon flux for each benchmark model is then calculated with
eq.~(\ref{eq-MuFluxSE}) using the branching ratios of
table~\ref{tab-BenchmarkBergstrom}.

The BMs have specific parameters combinations which are optimized to
give the right WMAP abundance, not specially a large flux of $Z$ and
by consequence not a large muon flux from the Earth (BM I has a
large branching ratio into a $Z$, but a relatively small
annihilation cross section). For this, we need, ideally, a large
branching ratio into $\gamma Z$ (like BM I and III) and a large
annihilation cross section (BM IV and, to lesser extent, BM III).
Additionally, a large cross-section via the Higgs is required (BM IV
and III again) so that capture in the Earth through SI interaction
is efficient, cf. eq.~(\ref{eq-sigmaH0NSI}) and
figure~\ref{fey-H0N-h-H0N}. Finally, we want a dark matter mass
close to the iron resonance. BM III realizes the best compromise,
but even in this case the flux is below the sensitivity of IceCube.
The signature in gamma rays from the GC, as envisioned
in~\cite{Gustafsson:2007pc}, may benefit from the possibility of
introducing a boost factor (BF), taking into account the uncertainty
regarding the abundance of dark matter at the GC. Although, with a
bit of luck, a signal in gamma rays could be seen even without a BF,
with BF$=10^2-10^4$, the signature would be spectacular. Such a BF
may be possible at the GC but is unlikely in the Solar system
vicinity, and {\em a fortiori} in the Earth.

\begin{table}[ht!]
\begin{center}
\begin{tabular}{c|l|c c c c}
\multicolumn{2}{r}{Benchmark Model} & \textbf{I} & \textbf{II} & \textbf{III} & \textbf{IV}  \\
\hline
& $m_h$ & 500 & 500 & 200 & 120\\
& $m_{H_0}$ & 70 & 50 & 70 & 70\\
Parameters & $m_{A_0}$ & 76 & 58.5 & 80 & 80\\
$[ \GeV \ ]$ & $m_{H^+}$ & 190 & 170 & 120 & 120\\
& $\mu_2$  & 120 & 120 & 125 & 95\\
& $\lambda_2 * 1 \GeV$ & 0.1 & 0.1 & 0.1 & 0.1\\
\hline
& $\Omega h^2$  & 0.1 & 0.1 & 0.12 & 0.11\\
& $\sigma v_{tot} \ [\percubcm \pers]$  & 1.6e-28 & 8.2e-29 & 8.7e-27 & 1.9e-26\\
\hline
& $\gamma$ Z & 33 & 0.6 & 2 & 0.1\\
Branching & $\gamma \gamma$ & 36 & 29 & 2 & 0.04\\
Ratio & $b \overline{b}$ & 26 & 60 & 81 & 85\\
$[ \ \% \ ]$ & $c \overline{c}$ & 2 & 4 & 5 & 5\\
& $\tau^+ \tau^-$ & 3 & 7 & 9 & 10 \\
\hline \hline
\textbf{Muon rate} & $\gamma-Z$ only & 4.12e-06 & 1.82e-06 & 3.36e-02 & 4.75e-03 \\
$[ \ \persqkm \peryr \ ]$ & $\gamma-Z$ + tree level  & 4.81e-06 & 2.63e-05 & 3.16e-01 & 8.58e-01\\
\end{tabular}
\end{center}
\caption[Parameters, branching ratios and muon flux for benchmark
models]{Parameters and branching ratios for the four IDM benchmark
models presented in~\cite{Gustafsson:2007pc}. The maximum muon
fluxes assuming no detector thresholds are far below the sensitivity
of IceCube. \label{tab-BenchmarkBergstrom}}
\end{table}

\subsubsection{Another non-minimal scenario: introducing Majorana neutrinos} \label{sec-Majorana}

\begin{figure}[htb!]
\begin{center}
\includegraphics[clip = true, viewport = 8.5cm 10.9cm 12.5cm 14.4cm, width=2.6cm]{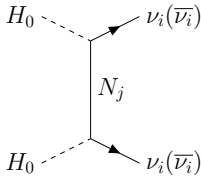}
\caption[Feynman diagram for direct annihilation of $H_0$ into
neutrinos with a Majorana neutrino in the propagator]{Direct
annihilation of $H_0$ into neutrinos $\nu_i$ (or antineutrinos
$\overline{\nu}_i$) through exchange of a Majorana neutrino $N_j$
leading to a mono-energetic neutrino flux. \label{fey-Majorana}}
\end{center}
\end{figure}

A natural extension of the IDM model consists in the addition of
right-handed Majorana neutrinos, odd under $Z_2$~\cite{Ma:2006km}.
This introduces the possibility to give a mass to the SM neutrinos
through radiative corrections. The IDM Lagrangian then contains
additional terms,
\begin{equation}
\mathcal{L} \supset  \, h_{ij} \, ( \nu_i \, H_0 - l_i \, H^+ )N_j +
{1\over 2} M_j  N_j N_j + h.c.\, \label{eq-MajYukawa}
\end{equation}
where the index $i$ stands for the SM neutrino generation ($i = e,
\mu, \tau$) and $j$ for the three extra Majorana neutrinos ($j =
1,2,3$). Further discussion about the generation of neutrino masses
can be found in~\cite{Ma:2006km}. In this extension of the IDM, the
dark matter candidate may be either a boson, the lighter of the two
neutral components of the $H_2$ doublet ($H_0$ or $A_0$), or a
fermion, if one of the Majorana neutrinos is the lightest odd
particle. As in the previous sections, we take the $H_0$ as dark
matter candidate and focus on the fact that the direct coupling
between $H_0$ and SM neutrinos allows annihilations of $H_0$ into
pairs of SM neutrinos or antineutrinos, with $E_{\nu,\bar\nu} =
m_{H_0}$, see figure~\ref{fey-Majorana}. In principle, annihilation
into $\nu\bar\nu$ is also possible. However, this process is either
p-wave suppressed, $\propto v_{\mathrm{rel}}^2$, or  helicity
suppressed, $\propto m_{\nu_i}^2$. The cross section for
annihilation into $\nu\nu$ or the conjugate channel,
$\bar\nu\bar\nu$, is of the form
\begin{equation}
\sigma |\vec{v}| ~ = ~ \frac{h^4}{4 \pi} ~ \frac{m_N^2}{(m_{H_0}^2 +
m_N^2)^2}\,. \label{eq-sigmaH0Majnu}
\end{equation}
To simplify our, exploratory level, discussion we focus on muonic
neutrinos and suppose that only one Majorana neutrino, with mass
$m_N$, enters the cross section. We do not consider the possibility
of oscillations, which is a good approximation for the Earth. For
reference, from here on we take $m_N = 100 \GeV$ and set the Yukawa
coupling to $h = 0.1$. These values are chosen so as to boost the
signal into mono-energetic neutrinos, which requires a rather light
Majorana neutrino. At the same time the SI cross section has to be
large enough to ensure efficient capture in the Earth. Meeting all
these constraints requires some fine tuning, but is possible for a
broad range of parameters.

\begin{figure}[htb!]
\begin{center}
\includegraphics[clip = true, viewport = 2.1cm 13.3cm 17.1cm 25.6cm, width=16cm]{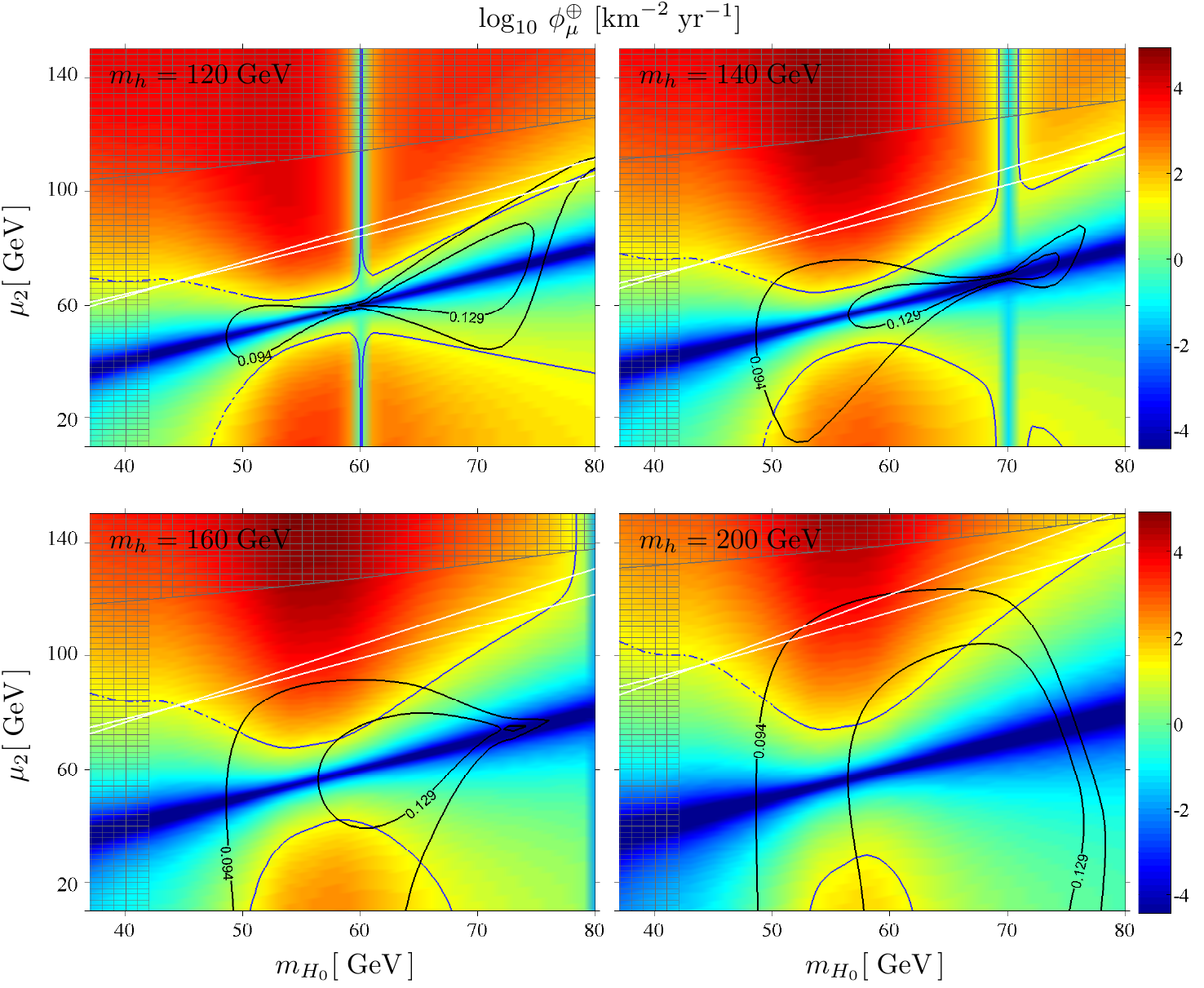}
\caption[Muon Flux from the Earth if the model is extended by three
Majorana neutrinos]{Muon flux resulting from mono-energetic
neutrinos from the Earth for the IDM including light
Majorana neutrinos.\\
\textsl{Colour gradient - $\, \log_{10} \phi_\mu ~ [\textsl{km}^{-2}
\, \textsl{yr}^{-1}]$; WMAP area (black lines); XENON, CDMS
exclusion limits (white lines); IceCube sensitivity hard
(blue lines); excluded regions (shaded). \\
(Parameters: $\lambda_2 = 0.2$, $\Delta m_{A_0 H_0} = 8 \,
\textsl{GeV}$, $\Delta m_{H^+ H_0} = 50 \, \textsl{GeV}$, $m_N = 100
\, \textsl{GeV}$, $h = 0.1$, $f = 0.3$).} \label{fig-MuFluxE-Maj}}
\end{center}
\end{figure}

The total muon flux coming only from annihilations into
mono-energetic neutrinos is shown in figure~\ref{fig-MuFluxE-Maj},
for different Higgs masses.
The plotted IceCube sensitivity curves (blue lines) are the ones for
a hard neutrino spectrum. The abundance used for the WMAP limit
(black lines) has been computed with \textsc{micrOMEGAs}. Note that
the shape of the WMAP regions differs from the ones of
figure~\ref{fig-MuFluxE-bm} as annihilations into neutrinos or
antineutrinos are also relevant at freeze-out.

For all Higgs masses but $m_h = 120 \GeV$, there is a region of the
parameter space that is in agreement with WMAP and for which the
muon flux is above the IceCube sensitivity. As in the previous
sections, uncertainties in $f$ affect the muon flux. We have set
$f=0.3$ but spanning over the range of $f$ preserves the overlap
observed, for example, for  $m_h = 200 \GeV$ while an increase of
$f$ for $m_h = 120 \GeV$ brings the WMAP region within the IceCube
sensitivity curve.

Figure~\ref{fig-MuFluxE-Maj-ScatPlot}, which is a scattered plot
over parameters of the model, gives  the total muon flux from the
Earth in function of the DM mass. The left panel shows that for the
IDM several points above the IceCube sensitivity are in agreement
with both WMAP and direct detection. Additionally, for the sake of
comparison, the muon fluxes for the MSSM are shown in the right
panel. The interesting IDM candidates are peaked around the iron
resonance ($50 \GeV < m_{H_0} < 60 \GeV$). For parameters in
agreement with WMAP (dark blue points for the IDM, yellow ones for
the MSSM), the IDM fluxes are typically about 2 orders of magnitude
larger than the MSSM ones and well above the IceCube sensitivity.
Actually, there are even a few points that are within the
sensitivity of Super-Kamiokande.

\begin{figure}[htb!]
\begin{center}
\includegraphics[clip = true, viewport = 2.1cm 5cm 17.1cm 12.6cm, width=16cm]{Figures/Earth_Maj.pdf}
\caption[Scattered Plot for the IDM with Majorana Extension and the
MSSM]{Scattered Plots of the muon fluxes of IDM and MSSM from the
Earth. IDM fluxes result from mono-energetic neutrinos from
the addition of light Majorana neutrinos. \\
\textit{left:} The blue points (upward triangles) correspond to
combinations of parameters that are in agreement with $2\sigma$ WMAP
and direct detection limits, the light blue ones (squares) are in
agreement with $2\sigma$ WMAP but the cross section is above the
direct detection limit, the orange ones (downward triangles) are
outside the $2\sigma$ WMAP region but the cross section is in
agreement with direct detection, the magenta ones (crosses) violate
both constraints. The grey points (dots) are excluded by other
constraints (e.g. vacuum stability). The red line gives the
IceCube sensitivity for a hard neutrino spectrum.\\
\textit{right:} All plotted points are within the $3\sigma$ WMAP
region. The blue ones correspond to the IDM with Majorana extension
(dark blue below, light blue above direct detection limit). The
green and yellow points are predicted by the MSSM (yellow below,
green above direct detection limit). The lines are different
detector thresholds: AMANDA-B10 (1997-1999) hard (solid) and soft
(dashed); AMANDA-II (2001-2003) hard (solid) and soft (dashed);
IceCube best case hard (solid) and soft (dashed); BAIKAL
(1998-2001); BAKSAN (1978-1995); MACRO (1989-1998) hard; SUPER-K
(1996-2001) soft, see ref.~\cite{Hubert:2007zza} and references therein. \\
\textsl{(Parameters: $m_h = 200 \, \textsl{GeV}$, $\lambda_2 = 0.2$,
$\Delta m_{A_0 H_0} = 8 \, \textsl{GeV}$, $\Delta m_{H^+ H_0} = 50
\, \textsl{GeV}$, $f = 0.3$, $m_N = 100 \, \textsl{GeV}$, $h =
0.1$).}\label{fig-MuFluxE-Maj-ScatPlot}}
\end{center}
\end{figure}
On the negative side, we have to face the fact that a boost of the
signal into mono-energetic neutrinos requires a rather light
Majorana particle ($m_N \sim 100 \GeV$) and a coupling to SM
neutrinos of the order $h \sim 0.1$. For those values of the
parameters, similar to those used in~\cite{Kubo:2006yx}, the
one-loop contribution to the masses of the SM neutrinos is quite
large, $m_{\nu} \sim \keV$ range, which clearly contradicts the
current limits on neutrino masses~\cite{Amsler:2008zzb}. One way
around is to make use of the fact that the absolute sign of fermions
is non-observable and to assume that there is another contribution
to the SM neutrino masses, beyond the IDM (say through
non-renormalisable operators) that compensates the contribution from
the light Majorana neutrino. Albeit this is contrived in our case,
as it would require a fine tuning of parameters at least at level
$10^{-3}$, a similar mechanism is sometimes invoked, for instance in
supersymmetric models with a sneutrino as the dark matter
candidate~\cite{Arina:2007tm}. Presumably, we should also make sure
that no other dangerous radiative processes are important, like $\mu
\rightarrow e + \gamma$~\cite{Kubo:2006yx}. This is, however, beyond
the scope of the present work.

\subsection{High mass candidate: indirect detection from the Galactic centre}

For the dark matter mass range under consideration in this section
($500 \GeV \lesssim m_{H_0} \lesssim 1500 \GeV$),\footnote{Higher
masses are in principle possible, but the range given here is
conservative, in the sense that we do not considered large quartic
couplings~\cite{LopezHonorez:2006gr}.} the most promising signal of
annihilation products comes from the Galactic centre. Earth and Sun
can not produce large muon fluxes since the cross section of dark
matter on nuclei, which is inversely proportional to the square of
the mass of the candidate, is too small.

In the framework of the basic IDM  and for this mass range,
annihilation into pairs of $W$ or $Z$ gauge bosons, or into Higgs
boson pairs, are dominant at tree level. Annihilation into
fermion-antifermion pairs through a Higgs boson, especially into $t
\bar t$, is also possible. However, for the parameter range
considered here, these channels turn out to be negligible.

The Higgs boson and the gauge bosons give distinct neutrino spectra.
Those originating from $W$ and $Z$ are considered as hard neutrinos:
they have a flat spectrum and a significant fraction has high
energies. Those produced through the decay of the Higgs are
comparatively softer. The amount of hard neutrinos among the total
neutrino flux is primarily determined by the branching ratio into
$W^+W^-$ and $ZZ$.

In ref.~\cite{LopezHonorez:2006gr}, it has been shown that, for the
mass range $500 \GeV \lesssim m_{H_0} \lesssim 1500 \GeV$, the
abundance is in agreement with WMAP for  $m_{H_0} \sim  \mu_2$.
Furthermore, all but a thin band around the diagonal is excluded by
vacuum stability and perturbativity. In the following, we consider
$\mu_2$ in the range $m_{H_0} - 20 \GeV \lesssim \mu_2 \lesssim
m_{H_0} + 15 \GeV$ and parameterize the $m_{H_0}$-$\mu_2$ plane by
($m_{H_0}; \, \mu_2-m_{H_0}$). The diagonal in the ($m_{H_0}$;
$\mu_2$) representation becomes the horizontal $\mu_2-m_{H_0} = 0$.
For the Higgs mass and the mass splittings, we follow the choice
made in the same reference,  $m_h = 120 \GeV$, $\Delta m_{A_0 H_0} =
5 \GeV$ and $\Delta m_{H^+ H_0} = 10 \GeV$. The coupling $\lambda_2
= 0.2$ chosen as in the previous sections completes the parameters
set.

\begin{figure}[htb!]
\begin{center}
\includegraphics[clip = true, viewport = 6.7cm 20.5cm 14.2cm 26.7cm, width=9.5cm]{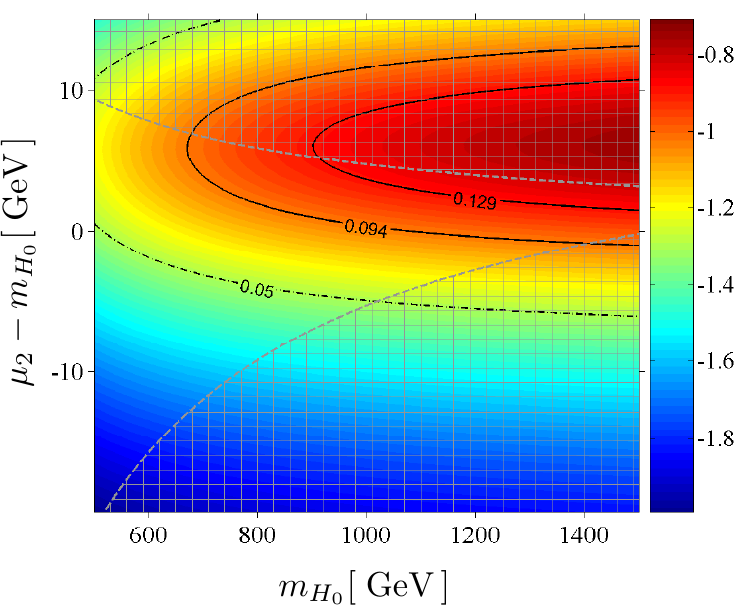}
\end{center}
\caption[Abundance GC]{Dark matter abundance obtained from
\textsc{micrOMEGAs}. The black lines represent the WMAP density
range at $2 \sigma$ (solid lines) and $3\sigma$ (dash-dotted line).
The shaded regions are excluded by vacuum
stability (upper part) and by perturbativity (lower part). \\
\textsl{Colour gradient - $\, \log_{10} \Omega h^2$; WMAP
area (black lines); excluded regions (shaded). \\
(Parameters: $m_h = 120 \, \textsl{GeV}$,  $\lambda_2 = 0.2$,
$\Delta m_{A_0 H_0} = 5 \, \textsl{GeV}$, $\Delta m_{H^+ H_0} = 10
\, \textsl{GeV}$).} \label{fig-AbundGC}}
\end{figure}

The abundance, computed with \textsc{micrOMEGAs}, is shown in
figure~\ref{fig-AbundGC}. The WMAP region is drawn in black, solid
(dash-dottet) lines corresponding to the $2\sigma$ (resp. $3\sigma$)
range. The exclusion limits (shaded region), as well as the WMAP
area, lie, as expected, close to the horizonal line, $\mu_2 -
m_{H_0} = 0$. The limit from vacuum stability corresponding to
$\lambda_2 = 0.2$ excludes the region $\mu_2 > m_{H_0}$. The
exclusion limit below $\mu_2 = m_{H_0}$ results from the less strict
perturbativity constraint, $1 < \lambda_i < 4\pi$.

The cross sections for the different annihilation channels, the
total annihilation cross section, and the branching ratios are also
computed with \textsc{micrOMEGAs} for the given choice of
parameters. It is found that, over the biggest part of the allowed
region, the $W^+ W^-$ channel is dominant, followed by $Z \, Z$. The
Higgs channel is only relevant in a region that is excluded by the
constraints mentioned above, while $t\overline{t}$ is negligible
everywhere. This implies that the neutrino flux is mainly hard for
the mass range considered in this section.

The total flux of neutrinos from the Galactic centre is computed
with eq.~(\ref{eq-diffFluxGC}), for the chosen combinations of
parameters. For this purpose, the integrated spectrum of neutrinos
({\em i.e.} the total number of neutrinos produced from all
annihilation channels) is obtained from \textsc{micrOMEGAs}. This
integration can be carried out for different energy thresholds. More
precisely, we do not only compute the total number of muonic, but
also that of electronic and tauonic neutrinos. Given the distance
from the Galactic centre, after neutrino oscillations, the three
generations arrive at the Earth with a ratio $\nu_\mu \, : \,
\nu_\tau \, : \,\nu_e \, = \, 1 \, : \, 1 \, : \,1$, and we obtain
the flux of muon-neutrinos by averaging over all generations. In
addition, we use the total annihilation cross section and
$\overline{J}$ taken from section~\ref{sec-NuFluxGC} for a NFW
profile and the opening angle of the ANTARES detector, a neutrino
telescope that may look at the Galactic centre.

The resulting neutrino flux is shown in figure~\ref{fig-NuFluxGC},
assuming a detector threshold of 30 GeV, characteristic of that
expected for ANTARES. In the plot on the left side, the colour
gradient is representing $\log_{10} \phi_\nu \ [\persqkm \peryr \,]$
and the black lines correspond to the $2\sigma$ WMAP limit. The
scattered plot on the right side gives  the neutrino flux only for
models that are not excluded by vacuum stability or perturbativity.
The colour code corresponds to whether the abundance is in agreement
with WMAP (dark blue) or not (light blue). The two red lines
represent the sensitivity of the ANTARES detector found in
ref.~\cite{Bailey2002} for the two extremals of a purely hard (solid
line) and a purely soft (dashed line) neutrino spectrum.

\begin{figure}[htb!]
\begin{center}
\includegraphics[clip = true, viewport = 3.2cm 12.5cm 17.7cm 18.5cm, width=16cm]{Figures/GC.pdf}
\end{center}
\caption[Neutrino Flux GC]{Expected neutrino flux in the ANTARES
detector from the Galactic centre, for a NFW profile, and a threshold energy of $30\GeV$.\\
\textit{left}: The colour gradient represents $\log_{10} \phi_\nu
[\persqkm \peryr]$. The WMAP region is drawn as black lines and the
excluded areas are shaded.\\
\textit{right}: Scattered plot of the neutrino flux together with
the sensitivity of ANTARES for a soft (dashed) and a hard (solid)
spectrum. The dark blue points lie within the WMAP area while the
light blue ones are outside. \\
\textsl{(Parameters: $m_h = 120 \, \textsl{GeV}$,  $\lambda_2 =
0.2$, $\Delta m_{A_0 H_0} = 5 \, \textsl{GeV}$, $\Delta m_{H^+ H_0}
= 10 \, \textsl{GeV}$).} \label{fig-NuFluxGC}}
\end{figure}

As in the non-excluded area, $H_0$ annihilates dominantly into
$W^+W^-$ and $Z Z$ gauge bosons, with a branching ratio $\gtrsim
80\%$, the neutrino flux is mainly hard and has to be compared to
the solid sensitivity curve. However, the expected neutrino flux
lies between two and three orders of magnitude below the expected
sensitivity of ANTARES.

\begin{figure}[htb!]
\begin{center}
\includegraphics[clip = true, viewport = 3.2cm 4.5cm 17.7cm 10.5cm, width=16cm]{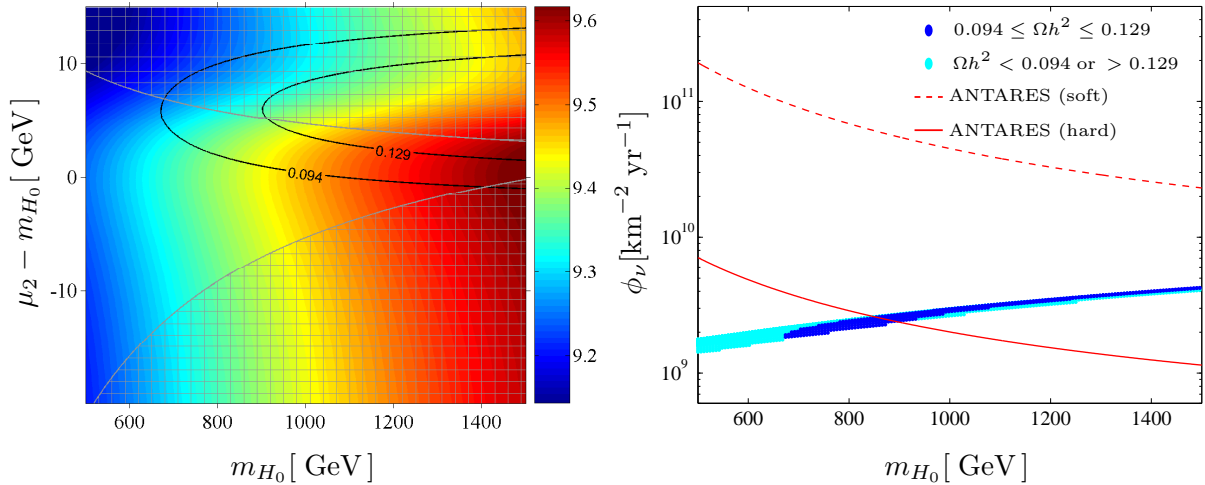}
\end{center}
\caption[Normalized Neutrino Flux GC]{Upper limit on the expected
neutrino flux from the IDM in the ANTARES detector. The neutrino
flux of figure~\ref{fig-NuFluxGC} is enhanced by a boostfactor
factor in order to circumvent the astrophysical uncertainties on the
dark matter profile at the Galactic centre and to obtain an upper
limit on the expectable neutrino flux while leaving the
corresponding
$\gamma$-flux below the EGRET limit.\\
\textit{left}: The colour gradient represents $\log_{10}
\phi_\nu^{norm} [\persqkm \peryr]$. The WMAP region is drawn as
black lines and the excluded areas are shaded.\\
\textit{right}: Scattered plot of the neutrino flux together with
the sensitivity of ANTARES for a soft (dashed) and a hard (solid)
spectrum. The dark blue points lie within the WMAP area
while the light blue ones are outside. \\
\textsl{(Parameters: $m_h = 120 \, \textsl{GeV}$,  $\lambda_2 =
0.2$, $\Delta m_{A_0 H_0} = 5 \, \textsl{GeV}$, $\Delta m_{H^+ H_0}
= 10 \, \textsl{GeV}$).} \label{fig-NuFluxNormGC}}
\end{figure}

As is usual in this discussions of indirect detection from the
Galactic centre, we make use of the fact that the dark matter
density profile in the innermost region of the Galaxy is poorly
known and might be steeper than the assumed NFW profile and enhanced
by some boost factor.  For instance, in~\cite{Gustafsson:2007pc},
the predicted gamma flux is boosted by a factor $\sim 10^4$
reflecting those uncertainties. The boost may also be partly of
particle physics origin. The annihilation of heavy dark matter
candidates that interact through relatively lighter particles (for
instance, the $W$, the $Z$ and the Higgs in the present case) may
have a larger cross section for low relative velocities than that
computed at tree level, a phenomenon called ''Sommerfeld
enhancement''~\cite{Hisano:2004ds,Cirelli:2007xd}. From the
similarity of the IDM with Minimal Dark Matter models, the expected
Sommerfeld enhancement in $H_0$ annihilations at the Galactic centre
is ${\cal O}(10^2)$ at most but might be non-negligible. Regardless
of the possibility of boosting the flux, we must ensure that the
signatures are not in contradiction with existing observations, say
in gamma rays, see for instance ref.~\cite{bertone-2004-70}. The
observations of the EGRET satellite of the Galactic centre, which we
take from the left-middle plot of figure~5c in
ref.~\cite{Hunter1997}, set an upper limit on the gamma rays that
may be produced in annihilations of the $H_0$. Similarly to the
computation of the neutrino spectrum, we have used
\textsc{micrOMEGAs} to obtain the differential photon spectrum,
together with $\overline{J}$ for a NFW profile and an opening angle
$\Delta \Omega =  10^{-3}$ corresponding to the resolution of  EGRET
(cf. section~\ref{sec-NuFluxGC}). For each point in the space of
models, we may compute the gamma ray flux and compare it to EGRET
data to derive a maximal boost factor (for the masses considered,
the most constraining observations are  at an energy $E_\gamma
\approx 10 \GeV$). Typically the maximal boost factors are $\sim
10^2 - 10^3$, which are actually quite moderate values, given the
uncertainties on the Galactic abundance and the possibility of
Sommerfeld enhancement. The maximum allowed neutrino flux is
obtained according to
\begin{equation}
\phi_\nu^{\mathrm{norm}} ~ = ~ \phi_\nu^{\mathrm{NFW}} \,
\frac{\phi_\gamma^{\mathrm{EGRET}}(E_{\mathrm{norm}})}{\phi_\gamma^{\mathrm{NFW}}(E_{\mathrm{norm}})}
\, ,
\end{equation}
for each model. The result is  shown in
figure~\ref{fig-NuFluxNormGC}. In the left plot, the WMAP area is
drawn in black and the excluded areas are shaded. The scattered plot
on the right contains only points that are not excluded, together
with the soft (given for information) and hard ANTARES sensitivities
in red. As discussed above, the neutrino flux should be compared to
the hard sensitivity (solid line),  due to the large branching ratio
into the $W^+ W^-$ and $Z Z$ channels. The dark blue colour
corresponds to models that agree within $2 \sigma$ with WMAP, the
light blue ones have too high or too low abundance.

For $H_0$ candidates with mass $\gtrsim 900 \GeV$, the maximum muon
flux is above the ANTARES sensitivity and  several models have an
abundance that is in agreement with WMAP. Similar predictions for
the neutralino may be found in ref.~\cite{bertone-2004-70}.

\section{Conclusion}

In this paper, we have investigated the indirect detection
signatures in neutrinos, for various scalar dark matter candidates
within the framework of the Inert Doublet Model. Concretely, we have
considered three distinct mass ranges. The first one corresponds to
a rather light WIMP, \mbox{$m_{H_0}~\sim$ few GeV}. This range is
not excluded by existing direct detection experiments, and may be
consistent with the combined results of DAMA/NaI and DAMA/Libra. We
have shown that the neutrinos that are produced by a light $H_0$
captured in the Sun can be constrained by Super-Kamiokande
measurements. This result corroborates the (model independent)
statements made in refs.~\cite{Feng:2008dz,Hooper:2008cf}. We have
also studied the predictions for more mundane WIMP candidates,
corresponding to $m_{H_0} \sim 60 \GeV$. Since the $H_0$ has only a
SI cross section on nuclei, the best hope to detect its presence is
in the Earth, in particular using the resonant scattering on iron.
Such signatures have also been considered in~\cite{Agrawal:2008xz}
and our results for the fluxes are in agreement. However, direct
detection strongly limits the potential of signatures from the Earth
(see for instance ref.~\cite{Barger:2007hj}), and the $H_0$ is no
exception. In particular, we have found no solutions that give an
observable signal from the Earth and are both in agreement with the
WMAP abundance and direct detection exclusion limits. In order to
boost the signal in neutrinos from the Earth, we have considered an
extension with (light) Majorana neutrinos. The model has to be
substantially fine-tuned to have WMAP abundance, capture in the
Earth and strong annihilations into mono-energetic neutrinos (as
well as reasonable SM neutrino masses), but some solutions may be
possible. These are depicted in
figure~\ref{fig-MuFluxE-Maj-ScatPlot}. Finally, we considered heavy
WIMPs, $m_{H_0} \sim $ TeV. Those candidates annihilate at the
centre of the Galaxy and produce high energy neutrinos that may be
observable, provided a boost factor $\sim 10^3$.

\section*{Note added in proof}

Recent simulations of the dark matter halo in presence of a disk of
baryons indicate that there might be a rotating disk of dark matter
in our Galaxy~\cite{Read:2008fh}. Because of the small relative
velocity between this so-called dark disk and the solar system, it
has been shown in~\cite{Bruch:2009rp} that capture in the Earth, and
to some extent in the Sun, might be dramatically enhanced. In that
paper, the impact of the dark disk on the neutrino and  muon fluxes
from the Earth is illustrated for CMSSM neutralino candidates, but
clearly similar enhancements should be expected for IDM dark matter
candidates ({\em i.e.} in particular for a candidate close to the
iron resonance, as considered in section~\ref{sec-middle}). We keep
for further work the investigation of this exciting possibility.

\section*{Acknowledgments}

This work is supported by the FNRS and Belgian Federal Science
Policy (IAP VI/11). The work of Q.S. and M.H.G.T is supported by the
Belgian Fonds de la Recherche Scientifique (FNRS). One of us (S.A.)
would like to thank Daan Hubert for discussions. We also thank
Joakim Edsj\"{o} for help on using \textsc{WimpSim} and for
enlightenment on capture in the Earth and Daniel Bertrand, Catherine
De Clercq, Jean-Marie Fr\`ere and Emmanuel Nezri for discussions.

\appendix

\bibliographystyle{utphys}
\bibliography{IDMref}
\addcontentsline{toc}{section}{References}

\end{document}